\documentclass[ALICE,manyauthors]{cernphprep}
\usepackage[comma,square,numbers,sort&compress]{natbib}
\usepackage{hyperref}
\usepackage{lineno}
\usepackage{xspace}
\usepackage{graphicx}
\usepackage[section]{placeins}
\usepackage[T1]{fontenc}
\usepackage{orcidlink}

\begin{document}
%

\newcommand{\pp}           {pp\xspace}
\newcommand{\ppbar}        {\mbox{$\mathrm {p\overline{p}}$}\xspace}
\newcommand{\XeXe}         {\mbox{Xe--Xe}\xspace}
\newcommand{\PbPb}         {\mbox{Pb--Pb}\xspace}
\newcommand{\pA}           {\mbox{pA}\xspace}
\newcommand{\pPb}          {\mbox{p--Pb}\xspace}
\newcommand{\AuAu}         {\mbox{Au--Au}\xspace}
\newcommand{\dAu}          {\mbox{d--Au}\xspace}

\newcommand{\s}            {\ensuremath{\sqrt{s}}\xspace}
\newcommand{\snn}          {\ensuremath{\sqrt{s_{\mathrm{NN}}}}\xspace}
\newcommand{\pt}           {\ensuremath{p_{\rm T}}\xspace}
\newcommand{\meanpt}       {$\langle p_{\mathrm{T}}\rangle$\xspace}
\newcommand{\ycms}         {\ensuremath{y_{\rm CMS}}\xspace}
\newcommand{\ylab}         {\ensuremath{y_{\rm lab}}\xspace}
\newcommand{\etarange}[1]  {\mbox{$\left | \eta \right |~<~#1$}}
\newcommand{\yrange}[1]    {\mbox{$\left | y \right |~<~#1$}}
\newcommand{\dndy}         {\ensuremath{\mathrm{d}N_\mathrm{ch}/\mathrm{d}y}\xspace}
\newcommand{\dndeta}       {\ensuremath{\mathrm{d}N_\mathrm{ch}/\mathrm{d}\eta}\xspace}
\newcommand{\avdndeta}     {\ensuremath{\langle\dndeta\rangle}\xspace}
\newcommand{\dNdy}         {\ensuremath{\mathrm{d}N_\mathrm{ch}/\mathrm{d}y}\xspace}
\newcommand{\Npart}        {\ensuremath{N_\mathrm{part}}\xspace}
\newcommand{\Ncoll}        {\ensuremath{N_\mathrm{coll}}\xspace}
\newcommand{\dEdx}         {\ensuremath{\textrm{d}E/\textrm{d}x}\xspace}
\newcommand{\RpPb}         {\ensuremath{R_{\rm pPb}}\xspace}

\newcommand{\nineH}        {$\sqrt{s}~=~0.9$~Te\kern-.1emV\xspace}
\newcommand{\seven}        {$\sqrt{s}~=~7$~Te\kern-.1emV\xspace}
\newcommand{\twoH}         {$\sqrt{s}~=~0.2$~Te\kern-.1emV\xspace}
\newcommand{\twosevensix}  {$\sqrt{s}~=~2.76$~Te\kern-.1emV\xspace}
\newcommand{\five}         {$\sqrt{s}~=~5.02$~Te\kern-.1emV\xspace}
\newcommand{\twosevensixnn}{$\sqrt{s_{\mathrm{NN}}}~=~2.76$~Te\kern-.1emV\xspace}
\newcommand{\fivenn}       {$\sqrt{s_{\mathrm{NN}}}=5.02$~Te\kern-.1emV\xspace}
\newcommand{\LT}           {L{\'e}vy-Tsallis\xspace}
\newcommand{\GeVc}         {Ge\kern-.1emV/$c$\xspace}
\newcommand{\MeVc}         {Me\kern-.1emV/$c$\xspace}
\newcommand{\TeV}          {Te\kern-.1emV\xspace}
\newcommand{\GeV}          {Ge\kern-.1emV\xspace}
\newcommand{\MeV}          {Me\kern-.1emV\xspace}
\newcommand{\GeVmass}      {Ge\kern-.2emV/$c^2$\xspace}
\newcommand{\MeVmass}      {Me\kern-.2emV/$c^2$\xspace}
\newcommand{\lumi}         {\ensuremath{\mathcal{L}}\xspace}

\newcommand{\ITS}          {\rm{ITS}\xspace}
\newcommand{\TOF}          {\rm{TOF}\xspace}
\newcommand{\ZDC}          {\rm{ZDC}\xspace}
\newcommand{\ZDCs}         {\rm{ZDCs}\xspace}
\newcommand{\ZNA}          {\rm{ZNA}\xspace}
\newcommand{\ZNC}          {\rm{ZNC}\xspace}
\newcommand{\SPD}          {\rm{SPD}\xspace}
\newcommand{\SDD}          {\rm{SDD}\xspace}
\newcommand{\SSD}          {\rm{SSD}\xspace}
\newcommand{\TPC}          {\rm{TPC}\xspace}
\newcommand{\TRD}          {\rm{TRD}\xspace}
\newcommand{\VZERO}        {\rm{V0}\xspace}
\newcommand{\VZEROA}       {\rm{V0A}\xspace}
\newcommand{\VZEROC}       {\rm{V0C}\xspace}
\newcommand{\Vdecay} 	   {\ensuremath{V^{0}}\xspace}

\newcommand{\ee}           {\ensuremath{e^{+}e^{-}}} 
\newcommand{\pip}          {\ensuremath{\pi^{+}}\xspace}
\newcommand{\pim}          {\ensuremath{\pi^{-}}\xspace}
\newcommand{\kap}          {\ensuremath{\rm{K}^{+}}\xspace}
\newcommand{\kam}          {\ensuremath{\rm{K}^{-}}\xspace}
\newcommand{\pbar}         {\ensuremath{\rm\overline{p}}\xspace}
\newcommand{\kzero}        {\ensuremath{{\rm K}^{0}_{\rm{S}}}\xspace}
\newcommand{\lmb}          {\ensuremath{\Lambda}\xspace}
\newcommand{\almb}         {\ensuremath{\overline{\Lambda}}\xspace}
\newcommand{\Om}           {\ensuremath{\Omega^-}\xspace}
\newcommand{\Mo}           {\ensuremath{\overline{\Omega}^+}\xspace}
\newcommand{\X}            {\ensuremath{\Xi^-}\xspace}
\newcommand{\Ix}           {\ensuremath{\overline{\Xi}^+}\xspace}
\newcommand{\Xis}          {\ensuremath{\Xi^{\pm}}\xspace}
\newcommand{\Oms}          {\ensuremath{\Omega^{\pm}}\xspace}
\newcommand{\Hee}          {\ensuremath{^{3}\mathrm{He}\xspace}}
\newcommand{\Heebar}       {\ensuremath{^{3}\overline{\mathrm{He}}\xspace}}
\newcommand{\degree}       {\ensuremath{^{\rm o}}\xspace}

\begin{titlepage}
\PHyear{2023}       
\PHnumber{268}      
\PHdate{20 November}  

\title{Measurements of chemical potentials in \PbPb collisions at \fivenn}
\ShortTitle{Measurements of chemical potentials in \PbPb collisions at \fivenn}   

\Collaboration{ALICE Collaboration\thanks{See Appendix~\ref{app:collab} for the list of collaboration members}}
\ShortAuthor{ALICE Collaboration} 

\begin{abstract}

This Letter presents the most precise measurement to date of the matter--antimatter imbalance at midrapidity in Pb--Pb collisions at a center-of-mass energy per nucleon pair \snn = 5.02 TeV. Using the Statistical Hadronization framework, it is possible to obtain the value of the electric charge and baryon chemical potentials, $\mu_Q=-0.18\pm0.90$~MeV and $\mu_B=0.71\pm0.45$~MeV, with unprecedented precision. A centrality-differential study of the antiparticle-to-particle yield ratios of charged pions, protons, $\Omega$-baryons, and light (hyper)nuclei is performed. These results indicate that the system created in Pb--Pb collisions at the LHC is on average baryon--free and electrically neutral at midrapidity.
\end{abstract}
\end{titlepage}

\setcounter{page}{2} 


\emph{Introduction.}
Nuclear matter at extremely high energy densities can be generated in the laboratory through relativistic heavy-ion collisions~\cite{Shuryak:2014zxa, Braun-Munzinger:2015hba, Busza:2018rrf}. At the LHC, the beam remnants from the collision are located at rapidities $y\approx\pm6$ and
a fraction of the collision energy is deposited at midrapidity~\cite{Mehtar-Tani:2009wji}. In this region, particles are formed from a nearly baryon number and electric charge free medium. This process can be described in the Color Glass Condensate model via gluon radiation by static quarks, frozen by time dilation~\cite{McLerran:1993ni}. Conversely, string-fragmentation models explain it through the breaking of color flux tubes. Part of the initial baryon number can be transported to midrapidity via either baryon junction formation~\cite{Vance:1997th} or diquark breaking~\cite{Capella:1996th}.
This phenomenon, known as nuclear stopping, influences the net-baryon density of the system formed at midrapidity~\cite{Videbaek:2005qn, 2007arXiv0705.2759T, Hoelck:2020iei}. The baryon number transport is minimal at the LHC, and the nuclear transparency regime~\cite{Elfner:2022iae} is reached. In this regime, conditions akin to those of the early Universe are replicated, where nearly equal abundances of matter and antimatter were present, as described by the standard cosmological model~\cite{Fromerth:2012fe}.
Experimentally, one can gauge the extent to which heavy-ion collisions approach the early Universe conditions by measuring the antimatter-to-matter yield ratios across various hadron species.

A comprehensive framework for interpreting these ratios is provided by the Statistical Hadronization Model (SHM)\cite{Braun-Munzinger:1998tqu, Cleymans:1998fq, Braun-Munzinger:2003pwq, Cleymans:2005xv, Braun-Munzinger:2008szb, Vovchenko:2016ebv}. Among the several models that can be used to describe a heavy-ion collision, the SHM is the most successful in describing the yields of all light-flavor hadronic species, which are determined starting from the partition function of the fireball at the freeze-out of inelastic scatterings. This fireball is an equilibrated gas composed of hadrons and resonances. Because of the substantial particle multiplicity and the finite kinematical acceptance, a Grand Canonical (GC) ensemble description is employed for heavy-ion collisions. In this approach, the conservation of charges, namely the baryon number ($B$), the electric charge ($Q$), and strangeness ($S$), is regulated by the corresponding chemical potentials $\mu_B$, $\mu_Q$, and $\mu_S$, respectively~\cite{Hagedorn:1984uy, Redlich:2001it}. The baryon chemical potential $\mu_B$ represents the net-baryon density of the system, with $\mu_B=0$ corresponding to an equilibrated gas composed of hadrons and resonances with same amount of baryons and antibaryons. The electric charge potential $\mu_Q$ encodes the positive-negative charge imbalance of the gas; it is connected to $\mu_B$ by the atomic-to-mass-number ratio $Z/A$ of the colliding ions~\cite{Bazavov:2012vg, Vovchenko:2019pjl}. The requirement of strangeness neutrality constrains $\mu_S$ throughout the entire volume of the fireball~\cite{Bazavov:2012vg, Vovchenko:2019pjl}. Chemical potentials determine the abundance of hadrons through the fugacity, $\lambda_i=\exp[(B_i\mu_B+Q_i\mu_Q+S_i\mu_S)/T_{\mathrm{ch}}]$, where $B_i$, $Q_i$, and $S_i$ denote the quantum numbers of the considered species $i$, and $T_\mathrm{ch}$ is the chemical freeze-out temperature, at which hadron yields are determined.

Over the last three decades, the asymmetry between antimatter and matter of the fireball has been systematically studied at different experimental facilities~\cite{Stachel:1989pa, Braun-Munzinger:1994ewq, E802:1998cxv,NA49:1999uyd, Braun-Munzinger:1999hun, Becattini:2000jw, Braun-Munzinger:2001uhy, STAR:2001rbj, BRAHMS:2001kqb, Braun-Munzinger:2001hwo, Magestro:2001jz, Becattini:2003wp, ALICE:2013mez, Andronic:2017pug, Andronic:2018qqt, Andronic:2021dkw}. The decreasing trend of $\mu_B$, from about $400\ \mathrm{MeV}$ at the SPS to $20\ \mathrm{MeV}$ at the top RHIC energy of 200 GeV, and $\mu_B=0.7\pm3.8\ \mathrm{MeV}$ at the LHC is consistent with the decrease of baryon number transport to midrapidity with increasing beam rapidity~\cite{NA49:1998nxh, PhysRevC.58.3523, E802:1999hit, E866:1999ktz, E877:1999qdc, AFANASIEV200059, E866:2000dog, E917:2000spt, E895:2001yfr, E895:2001zms, PhysRevLett.88.062301, STAR:2002hpr, STAR:2002npn, NA49:2002pzu, PHENIX:2002svd, STAR:2002ygs, NA44:2002wha, E917:2003gec, E-0895:2003oas, E895:2003qcm, PHENIX:2003wtu, PhysRevC.69.024902, NA49:2003bok, PHENIX:2003iij, STAR:2003jis, STAR:2003jwm, NA49:2004irs, NA57:2004nxc, STAR:2004bgh,PHENIX:2004spo, PHENIX:2004vqi, STAR:2004yym, Anja_2004, Zhang:2004rj, Barannikova:2004rp, NA57:2005ftu, BRAHMS:2005gow, Andronic:2005yp, Andronic:2008gu, Alba:2014eba, Andronic:2017pug, STAR:2017sal, Andronic:2018qqt}. The formation of baryon number free matter at midrapidity was first reported in pp collisions by ALICE, which observed that the $\bar{\mathrm{p}}/\mathrm{p}$ yield ratio is compatible with unity~\cite{ALICE:2010hjm}. At fixed collision energy, it is also possible to explore nuclear transparency as a function of centrality, i.e., the transverse displacement between the centers of the colliding nuclei, as it affects the dynamics of the colliding nucleons. In particular, a slight increase in $\mu_B$ from peripheral to central (head-on) collisions was observed at low energies by STAR at the RHIC beam energy scan~\cite{STAR:2017sal}. These results were obtained by either comparing the SHM predictions with the measured yields of hadrons and their antimatter counterparts~\cite{Cleymans:2006xj} or by directly fitting antiparticle-to-particle yield ratios~\cite{Andronic:2005yp, STAR:2017sal}.

In this Letter, we report the most precise estimation to date of $\mu_B$ and $\mu_Q$ obtained from a set of antiparticle-to-particle yield ratios.
Compared to previous estimations, the precision of the current results has improved by about an order of magnitude.
This improvement in precision is attributed to the proper treatment of the
cancelation of particle-antiparticle correlated uncertainties and the reduced dependence on model parameters, such as the system volume, $V$, which is eliminated in the antiparticle-to-particle yield ratios.
The analyzed species are charged pions, protons, $\Omega^-$ baryons, and light (hyper)nuclei. (Anti)protons are the most abundantly produced (anti)baryons at midrapidity ($\approx$~35 and $\approx$~2 protons on average in central and peripheral Pb--Pb collisions, respectively~\cite{ALICE:2019hno}). Consequently, the antiproton-to-proton yield ratio can probe the antibaryon-to-baryon imbalance~\cite{ALICE:2010hjm, ALICE:2013yba} with high precision. On the other hand, the sensitivity to baryon asymmetry is enhanced when light (hyper)nuclei are included because of their larger baryon content. In this work, $^3\mathrm{He}$, its isobar $^3\mathrm{H}$, and hypertriton $^3_{\Lambda}\mathrm{H}$, which is a bound state of a proton, a neutron, and a $\Lambda$, along with their antimatter counterparts, are considered\footnote{(Anti)deuterons, $\mathrm{d}$($\bar{\mathrm{d}}$) are not considered in this Letter since the efficiency correction for $\bar{\mathrm{d}}$ is based on the $\bar{\mathrm{d}}$ absorption cross section extracted by the ALICE Collaboration from the measured $\bar{\mathrm{d}}/\mathrm{d}$ yield ratio itself~\cite{ALICE:2020zhb}.}. The ratio of oppositely charged pions provides a precise constraint on the imbalance of electric charge, as the yield ratio depends predominantly on $\mu_Q$. Finally, the dependence of antimatter-to-matter ratios on strangeness is probed with (anti)$\Omega^-$ baryons, which, unlike (anti)$\Lambda$ and (anti)$\Xi^{-}$, have negligible contamination coming from heavier hadron decays.

\emph{The ALICE detector and data analysis}.
The results reported in this analysis are obtained from a sample of \PbPb collisions at \fivenn collected in 2018 by ALICE at the LHC. The ALICE apparatus and its performance are described in detail in~ Refs.~\cite{ALICE:2008ngc, ALICE:2014sbx}. The minimum-bias collision and centrality triggers are provided by the V0 system~\cite{ALICE:2013axi}, which is composed of two arrays of plastic scintillators covering the forward ($2.8<\eta<5.1$) and backward ($-3.7<\eta<-1.7$) regions of pseudorapidity. The coincidence of signals in both detectors determines the minimum bias trigger. The amplitude of the V0 signal is proportional to the charge deposited in the detectors, which is related to the produced charged-particle multiplicity that, in turn, is controlled by the collision centrality. The V0 amplitude is then used to trigger specific categories of central and semicentral events, and to estimate centrality~\cite{ALICE:2018tvk}. Five centrality intervals are considered in this Letter, namely 0--5\%, 5--10\%, 10--30\%, 30--50\%, and 50--90\%, expressed as percentiles of the total hadronic cross section for Pb--Pb collisions. The position of the primary interaction vertex is required to be within a 10 cm wide region centered at the nominal interaction point to profit from the full acceptance of the ALICE central barrel detectors. Events with multiple interaction vertices are rejected to ensure the correct association of reconstructed tracks and primary vertices. The number of events passing these selections is approximately 300 million.

Charged pions, protons, \Hee, and tritons produced at midrapidity, $\vert y\vert <0.5$, are tracked in the ALICE central barrel: hereafter, charge conjugates are implied unless stated otherwise. The tracks are reconstructed within $\vert\eta\vert<0.8$ and in the full azimuth using the Inner Tracking System (ITS)~\cite{ALICE:2010tia} and the Time Projection Chamber (TPC)~\cite{Alme:2010ke}. These detectors are placed in a solenoid that provides a uniform magnetic field of 0.5 T parallel to the beam axis. The antiparticle-to-particle yield ratios are measured as a function of the transverse momentum $p_{\mathrm{T}}$ in the ranges $0.7\leq p_{\mathrm{T}}<1.6\ \mathrm{GeV}/c$ for $\pim/\pip$, $0.5\leq p_{\mathrm{T}}<3\ \mathrm{GeV}/c$ for $\bar{\mathrm{p}}/\mathrm{p}$, $1.6\leq p_{\mathrm{T}} < 3\ \mathrm{GeV}/c$ for ${}^{3}\overline{\mathrm{H}}/^{3}\mathrm{H}$, and $2\leq p_\mathrm{T}<8\ \mathrm{GeV}/c$ for ${}^{3}\overline{\mathrm{He}}/^{3}\mathrm{He}$ to select the bulk of the production and ensure good identification performance.

The analysis procedure for extracting particle yields is similar to the one adopted in previous analyses~\cite{ALICE:2015wav, ALICE:2019hno, ALICE:2022veq}. Standard selections on the $\chi^{2}$ of the track fit, on the number of reconstructed track points in the ITS and the TPC, and on the distance of closest approach of the extrapolation of the track to the primary interaction vertex ensure a good reconstruction of tracks originating from the collisions. Particle identification is performed on a statistical basis by measuring the specific energy loss ($\mathrm{d}E/\mathrm{d}x$) in both the TPC and the ITS, and particle velocity depending on the transverse momentum of the measured particles with the Time-Of-Flight detector (TOF). Further details about the particle identification are provided in Appendix~\ref{app:pid}.

The residual contamination due to hyperon weak decays and spallation reactions of primary particles in the apparatus is evaluated by fitting the measured distance of closest approach distribution in the plane transverse to the beam axis with templates computed via Monte Carlo (MC) simulations for the various processes involved~\cite{ALICE:2015wav, ALICE:2019hno, ALICE:2022veq}. The extracted yields are corrected for the detector acceptance and candidate selection efficiency, computed using MC simulations, as the fraction of particles reconstructed out of all MC-generated primary particles. The Pb--Pb event is generated with HIJING~\cite{Wang:1991hta}, while the particles are transported through a realistic model of the ALICE apparatus with GEANT4~\cite{GEANT4:2002zbu}. To increase the simulated sample size protons, \Hee ~nuclei, and tritons are injected on top of each HIJING event. The available measurements of hadron inelastic cross sections are used to correct the GEANT4 parameterizations of the corresponding reactions~\cite{Chen1955CH67, Booth1957_exforO2097.1, Porile1962PO12, Artun1975AR17, Sadler1980SA13, Ashery:1981tq, Nakamura:1984xw, McGill1984MC02, Zaitseva1992ZA02, Kuzichev:1993tn, Herbach2006HE16, Zamani2010ZA05, Bardayan2010BA21, ALICE:2022zuz, ALICE:2023aod}.

The $^{3}_{\Lambda}\mathrm{H}$ candidates are reconstructed from their two-body charged mesonic decay $^{3}_{\Lambda}\mathrm{H}\to{}^{3}\mathrm{He}+\pi^-$. The reconstruction algorithm is the same as the one applied in previous measurements~\cite{ALICE:2015oer, ALICE:2019vlx, ALargeIonColliderExperiment:2021puh, ALICE:2022sco}. The $\Omega^-$ is reconstructed with a similar procedure from the decay into a charged kaon and a $\Lambda$ baryon, that, in turn, is reconstructed from its charged two-body decay, $\Omega^-\to \mathrm{K^- + \Lambda(\to \pi^- + p)}$~\cite{ALICE:2012yqk, ALICE:2013xmt, ALICE:2015mpp}. The ratios are extracted in intervals of proper decay length $ct=cML/p$, with $M$, $L$, and $p$ being the mass, trajectory length, and candidate momentum, respectively. In particular, $2\leq ct<35\ \mathrm{cm}$ for $^{3}_{\Lambda}\mathrm{H}$ and $1\leq ct<10\ \mathrm{cm}$ for $\Omega^-$ are used. The $^{3}_{\Lambda}\mathrm{H}$ and $\Omega^{-}$ candidates are selected with Boosted Decision Tree algorithms~\cite{10.1145/2939672.2939785}, which are applied on top of preliminary kinematic and topological selections to enhance the background rejection. The Boosted Decision Tree internal parameters and selections are optimized using samples of correctly classified signal and background candidates, as explained in detail in Appendix~\ref{app:ml}.

The invariant mass distribution of the selected candidates is fitted with a probability density function built with a Kernel Density Estimation~\cite{Cranmer:2000du, verkerke2003roofit} in the MC for $^{3}_{\Lambda}\mathrm{H}$, whereas an extended Crystal-Ball function is used for the $\Omega^{-}$ signal~\cite{ALICE:2015ikg}. An exponential function is used to model the residual background in both cases. The yields extracted as the integral of the signal functions obtained from the fits are corrected by the overall selection efficiency and acceptance computed in the MC simulations. As in previous $^{3}_{\Lambda}\mathrm{H}$ analyses~\cite{ALICE:2022sco}, an absorption correction factor is included to account for undetected candidates absorbed in the detector material before their decay.

The following systematic uncertainty contributions are estimated for the antiparticle-to-particle yield ratios: candidate selection and signal extraction, MC data sample size, material budget uncertainty, absorption cross section uncertainties, and magnetic field polarity. The details about the estimation and values of such contributions are reported in Appendix~\ref{app:sys}.

\emph{Results}.
The fully corrected antiparticle-to-particle yield ratios do not exhibit any significant dependence on $p_{\mathrm{T}}$ and $ct$ (see Appendix~\ref{app:ratiosResults}). This observation, which is consistent across particle species and centrality intervals, implies that the production spectra of charge-conjugate species only differ by normalization factors proportional to their yields. The antiparticle-to-particle yield ratios of each species are obtained as the averages weighted with the total uncorrelated uncertainties of the $p_{\mathrm{T}}$- and $ct$-differential ratios in each centrality interval. For $^{3}_{\Lambda}\mathrm{H}$, no statistically significant signal is observed in the 50--90\% centrality range.

\begin{figure}[tb]
    \begin{center}
    \includegraphics[width = 0.9\textwidth]{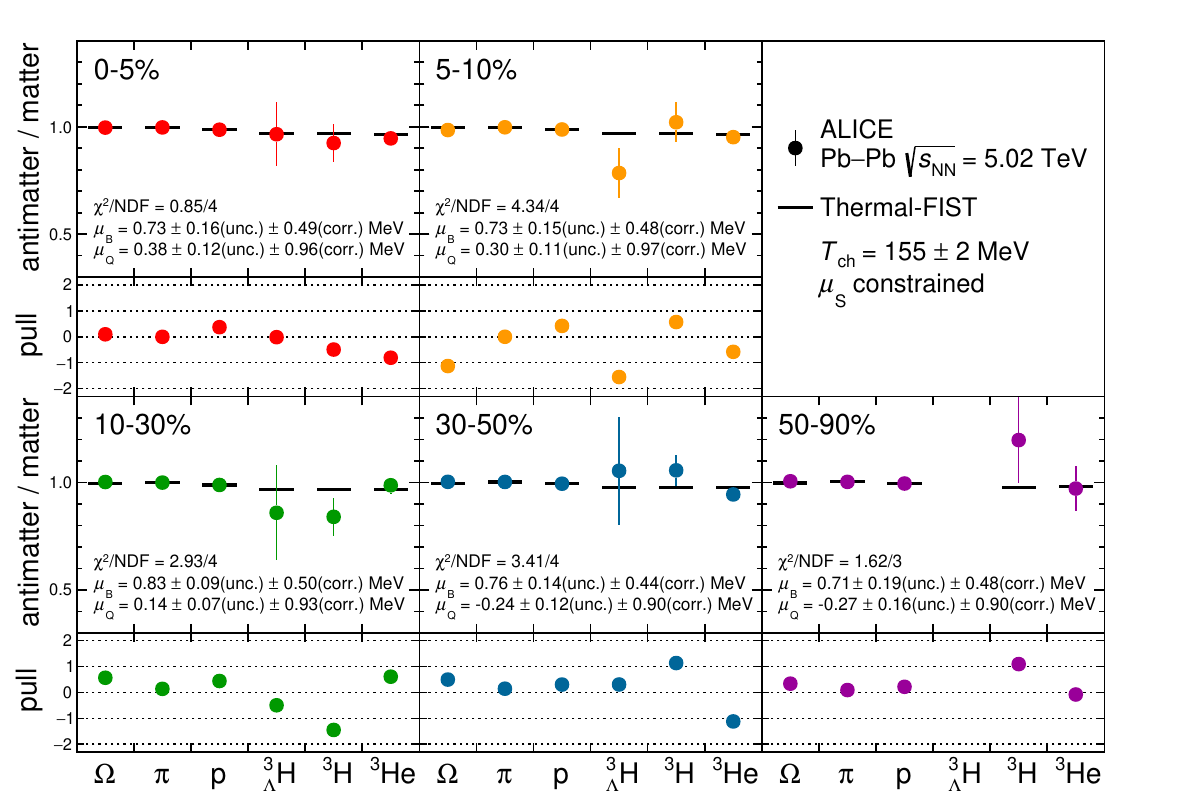}
    \end{center}
    \caption{Upper panels: statistical hadronization model fits to the measured antiparticle-to-particle yield ratios in different centrality intervals. Error bars show the sum in quadrature of statistical and centrality-uncorrelated systematic uncertainties. When not visible, error bars are hidden by the marker. Lower panels: pull distribution, defined as the difference between data and fit values, normalized to the uncertainty in the data.}
    \label{fig:TheFistFit}
\end{figure}

The chemical potentials $\mu_B$ and $\mu_Q$ are extracted by fitting the antiparticle-to-particle yield ratios with the predictions of the GC statistical hadronization model using the Thermal-FIST code~\cite{Vovchenko:2019pjl}. The measured ratios and the SHM fit results are reported in Fig.~\ref{fig:TheFistFit}. The chemical freeze-out temperature is set to $T_\mathrm{ch}=155\pm2\ \mathrm{MeV}$, as obtained from a fit to the ALICE data~\cite{Vovchenko:2018fmh, ALICE:2022wpn}: its value is compatible with the pseudo-critical temperature extracted with lattice QCD calculations~\cite{Borsanyi:2020fev}. This value is fixed for all centralities, since in heavy-ion collisions only a mild dependence of $T_{\mathrm{ch}}$ on centrality is observed (less than 3\%~\cite{STAR:2008med, ALICE:2013mez, STAR:2017sal, Vovchenko:2018fmh}); additionally, antiparticle-to-particle yield ratios show a negligible dependence on $T_{\mathrm{ch}}$ for $\mu_B\approx1\ \mathrm{MeV}$~\cite{Cleymans:2006xj}. The uncertainty on $T_{\mathrm{ch}}$, which is compatible with the range of variations of $T_{\mathrm{ch}}$ observed as a function of centrality, is considered as a centrality-correlated source of systematic uncertainty. The strangeness chemical potential $\mu_S$ is constrained in the fit from strangeness conservation. The contribution of strongly-decaying resonances is accounted in the model predictions as it cannot be directly disentangled in the data. For the $\chi^2$ minimization, the quadratic sum of statistical and uncorrelated systematic uncertainty is considered. The effect of the centrality-correlated sources is evaluated by repeating the fit to ratios coherently increased or decreased by their uncertainties. The uncertainty assigned to $\mu_B$ and $\mu_Q$ is half of the difference between the results obtained in the two cases.

In this Letter, yield ratios are analyzed within the GC statistical model also in the most peripheral events, where canonical ensemble formulation is needed for an accurate description of hadron yields by requiring exact conservation of charges over a finite volume~\cite{Vovchenko:2019kes, Cleymans:2020fsc}. It is known, however, that effects connected to the canonical conservation of charges cancel out when considering antiparticle-to-particle yield ratios, and their values are well described by the GC ensemble~\cite{Cleymans:2000ck, Braun-Munzinger:2003pwq}. Indeed, good fit quality is obtained across the 0--90$\%$ centrality range using the GC model to quantify these ratios.
In addition, the yield ratio $\overline{\Omega}^+/\Omega^-$ is compatible with unity as expected in the SHM, where it is weakly dependent of $\mu_B$ and $\mu_S$ for $\mu_B\sim0$~\cite{Cleymans:2005xv}.

\begin{figure}[t]
    \begin{center}
    \includegraphics[width=0.49\textwidth]{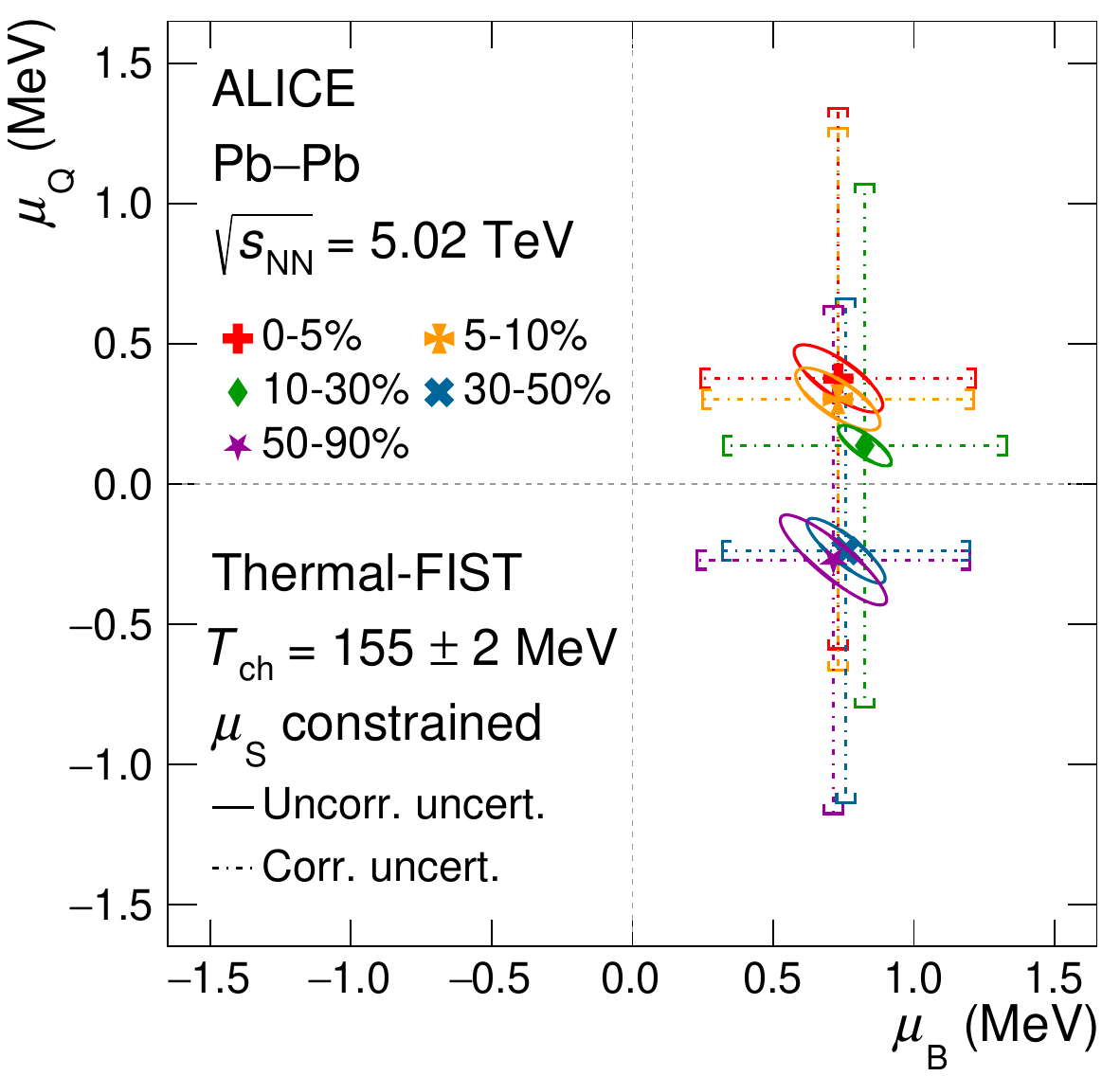}
    \includegraphics[width = 0.49\textwidth]{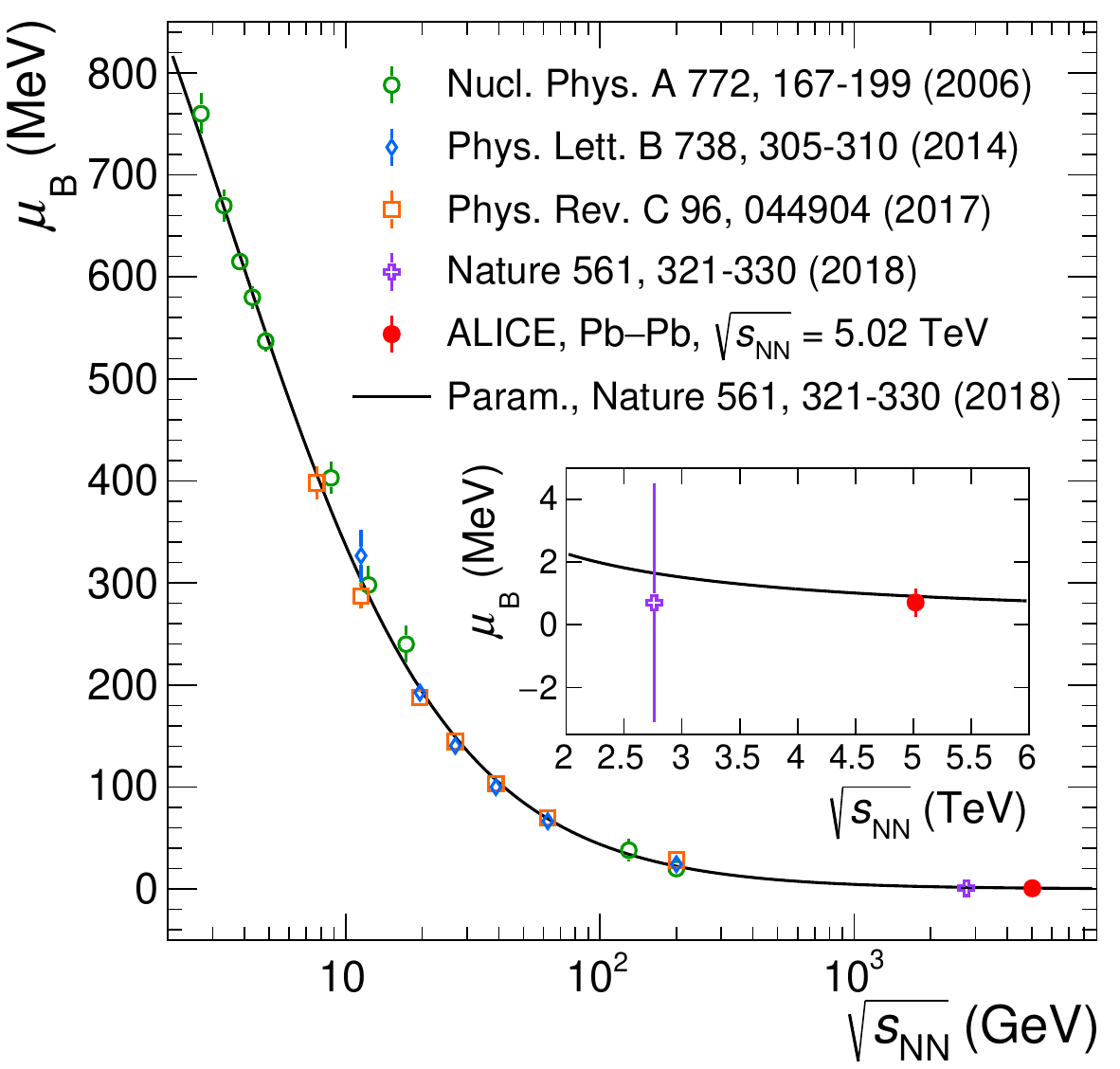}
    \end{center}
    \caption{Left panel: $\mu_B$ and $\mu_Q$ obtained with Thermal-FIST~\cite{Vovchenko:2019pjl} in different centrality intervals. The centrality-correlated and centrality-uncorrelated uncertainties are represented with error bars and ellipses, respectively. Right panel: $\mu_B$ extracted from data collected in Au--Au and Pb--Pb collisions at the AGS (E802, E866, E877, E895, E896, E917 Collaborations), SPS (NA44, NA49, NA47 Collaborations), RHIC (BRAHMS, PHENIX, STAR Collaboration), and LHC (ALICE Collaboration) as a function of the center-of-mass energy per nucleon--nucleon pair~\cite{Andronic:2005yp,Alba:2014eba,STAR:2017sal}, and phenomenological parameterization of $\mu_B(\sqrt{s_{\mathrm{NN}}})$~\cite{Andronic:2017pug}. The inset shows more in detail the results obtained at the LHC~\cite{Andronic:2017pug}.}
    \label{fig:MuQMuBandEnergyDependence}
\end{figure}

The chemical potentials obtained in different centrality intervals are shown in the left panel of Fig.~\ref{fig:MuQMuBandEnergyDependence}. The contours show a negative correlation between $\mu_B$ and $\mu_Q$, which is connected to the approximate exponential dependence of antiparticle-to-particle yield ratios on the linear combination of the chemical potentials. The centrality dependence of $\mu_B$ and $\mu_Q$ is studied by fitting independently the centrality-differential $\mu_B$ and $\mu_Q$ results with a constant function, taking into account the full correlation matrix of the measurements. Both the correlation matrices and the $\chi^{2}$ profiles of the fits are reported in the Appendix. The fit probability is $P=0.97$ for $\mu_B$ and $P=0.64$ for $\mu_Q$: therefore, no evidence of centrality dependence is found, even if a larger $\mu_B$ would be expected in more central collisions due to a potentially larger baryon stopping~\cite{Mehtar-Tani:2009wji}. The fit of the centrality-differential values yields chemical potentials $\mu_B=0.71\pm0.45\ \mathrm{MeV}$ and $\mu_Q=-0.18\pm0.90\ \mathrm{MeV}$, which are compatible with zero within $1.6\sigma$ and $0.2\sigma$, respectively. The comparison with the previous data point of $\mu_B$ at the LHC~\cite{ALICE:2013mez, Andronic:2017pug, Andronic:2018qqt, Andronic:2021dkw} shows a significant improvement in the precision by a factor larger than eight (no direct value of $\mu_Q$ was provided in that study, see below). These results imply that the system created at midrapidity in Pb--Pb collisions is baryon- and electrically-neutral on average. As a consequence, this observation shows that the nuclear transparency regime is reached, i.e., baryon transport from the colliding ions to the interaction region is negligible. Because of the absence of any centrality dependence, it is also concluded that nuclear transparency is achieved even in central \PbPb collisions, where a larger-than-zero $\mu_B$ could be expected from a more significant baryon number transport at midrapidity.

As a cross check, the SHM fits described above are repeated by also constraining $\mu_Q$ from initial conditions via conservation laws, as it was done also in past measurements~\cite{Andronic:2005yp, STAR:2017sal, Andronic:2017pug}. Specifically, the $\mu_Q/\mu_B$ ratio is fixed by requiring that the average charge-to-baryon density ratio of the created hadron system, $\langle n_Q\rangle/\langle n_B\rangle $, is equivalent to the $Z/A$ ratio of colliding nuclei, i.e., $\langle n_Q\rangle/\langle n_B\rangle=Z/A\approx0.4$ for ${}^{208}\mathrm{Pb}$~\cite{Bazavov:2012vg}. The $\mu_B$ values extracted from the fits in each centrality interval are successfully fitted with a constant function (fit probability $P=0.09$). The resulting $\mu_B$ value is compatible with the one reported above within uncertainties. Similar results are obtained by fitting the antiparticle-to-particle yield ratios using the GSI-Heidelberg model~\cite{Braun-Munzinger:2003pwq, Andronic:2005yp, Andronic:2018qqt}, with $T_{\mathrm{ch}}=156.6\pm1.7\ \mathrm{MeV}$~\cite{Andronic:2021dkw} and $\mu_Q$ is fixed to initial conditions: the average value across centrality is $\mu_B=0.90\pm0.43\ \mathrm{MeV}$. The $\chi^2$ profile of the fit is reported in Appendix~\ref{app:profiles}.
Using the values of $\mu_{B}$ and $\mu_Q$ extracted in the 5\% most central collisions, the inclusive net-proton density at midrapidity, $2/\langle N_{\mathrm{part}}\rangle \mathrm{d}N_{\mathrm{p} - \overline{\mathrm{p}}}/\mathrm{d}y$, can be computed in the SHM framework. The value extracted with Thermal-FIST is $(3.4 \pm 1.4)\times 10^{-3}$, while using the GSI-Heidelberg model, a value of $5.9^{+2.2}_{-2.8}\times10^{-3}$ is obtained. In both cases, the obtained results agree with the exponential trend as a function of beam rapidity predicted by the baryon-junction mechanism~\cite{Brandenburg:2022hrp}.

The right panel of Fig.~\ref{fig:MuQMuBandEnergyDependence} shows the comparison of the current with past estimations of $\mu_B$ as a function of the center-of-mass energy of the collision~\cite{Andronic:2005yp, Alba:2014eba, STAR:2017sal, Andronic:2017pug}. The comparison with the previous LHC data point is highlighted in the inset of the figure. The result reported in this Letter is compatible with the extrapolation of the phenomenological parameterization based on previous data and reported in Ref.~\cite{Andronic:2017pug}.

\emph{Conclusions}.
In summary, the most precise measurement of the asymmetry between matter and antimatter at the LHC is reported in this Letter. The asymmetry is quantified through antiparticle-to-particle yield ratios of different hadrons, which are analyzed within the statistical hadronization framework to extract the chemical potentials $\mu_B$ and $\mu_Q$. The GC version of the model accurately describes the antiparticle-to-particle yield ratios across centrality, indicating the elimination of effects from canonical charge conservation in peripheral events. The cancelation of correlated uncertainties in these ratios leads to a significant improvement in the $\mu_B$ precision: the uncertainty on the obtained value is about one order of magnitude smaller than the previously published one~\cite{Andronic:2017pug}. In addition, a direct estimation of $\mu_Q$ is provided. Furthermore, the first centrality-differential study of chemical potentials at the LHC is reported in this Letter. The obtained chemical potentials are consistent with zero, i.e., with the nuclear transparency regime being reached across the full centrality range, thus indicating that baryon transport to midrapidity is negligible even in the most central events at the LHC.


\newenvironment{acknowledgement}{\relax}{\relax}
\begin{acknowledgement}
\section*{Acknowledgements}

The ALICE Collaboration would like to thank all its engineers and technicians for their invaluable contributions to the construction of the experiment and the CERN accelerator teams for the outstanding performance of the LHC complex.
The ALICE Collaboration gratefully acknowledges the resources and support provided by all Grid centres and the Worldwide LHC Computing Grid (WLCG) collaboration.
The ALICE Collaboration acknowledges the following funding agencies for their support in building and running the ALICE detector:
A. I. Alikhanyan National Science Laboratory (Yerevan Physics Institute) Foundation (ANSL), State Committee of Science and World Federation of Scientists (WFS), Armenia;
Austrian Academy of Sciences, Austrian Science Fund (FWF): [M 2467-N36] and Nationalstiftung f\"{u}r Forschung, Technologie und Entwicklung, Austria;
Ministry of Communications and High Technologies, National Nuclear Research Center, Azerbaijan;
Conselho Nacional de Desenvolvimento Cient\'{\i}fico e Tecnol\'{o}gico (CNPq), Financiadora de Estudos e Projetos (Finep), Funda\c{c}\~{a}o de Amparo \`{a} Pesquisa do Estado de S\~{a}o Paulo (FAPESP) and Universidade Federal do Rio Grande do Sul (UFRGS), Brazil;
Bulgarian Ministry of Education and Science, within the National Roadmap for Research Infrastructures 2020-2027 (object CERN), Bulgaria;
Ministry of Education of China (MOEC) , Ministry of Science \& Technology of China (MSTC) and National Natural Science Foundation of China (NSFC), China;
Ministry of Science and Education and Croatian Science Foundation, Croatia;
Centro de Aplicaciones Tecnol\'{o}gicas y Desarrollo Nuclear (CEADEN), Cubaenerg\'{\i}a, Cuba;
Ministry of Education, Youth and Sports of the Czech Republic, Czech Republic;
The Danish Council for Independent Research | Natural Sciences, the VILLUM FONDEN and Danish National Research Foundation (DNRF), Denmark;
Helsinki Institute of Physics (HIP), Finland;
Commissariat \`{a} l'Energie Atomique (CEA) and Institut National de Physique Nucl\'{e}aire et de Physique des Particules (IN2P3) and Centre National de la Recherche Scientifique (CNRS), France;
Bundesministerium f\"{u}r Bildung und Forschung (BMBF) and GSI Helmholtzzentrum f\"{u}r Schwerionenforschung GmbH, Germany;
General Secretariat for Research and Technology, Ministry of Education, Research and Religions, Greece;
National Research, Development and Innovation Office, Hungary;
Department of Atomic Energy Government of India (DAE), Department of Science and Technology, Government of India (DST), University Grants Commission, Government of India (UGC) and Council of Scientific and Industrial Research (CSIR), India;
National Research and Innovation Agency - BRIN, Indonesia;
Istituto Nazionale di Fisica Nucleare (INFN), Italy;
Japanese Ministry of Education, Culture, Sports, Science and Technology (MEXT) and Japan Society for the Promotion of Science (JSPS) KAKENHI, Japan;
Consejo Nacional de Ciencia (CONACYT) y Tecnolog\'{i}a, through Fondo de Cooperaci\'{o}n Internacional en Ciencia y Tecnolog\'{i}a (FONCICYT) and Direcci\'{o}n General de Asuntos del Personal Academico (DGAPA), Mexico;
Nederlandse Organisatie voor Wetenschappelijk Onderzoek (NWO), Netherlands;
The Research Council of Norway, Norway;
Commission on Science and Technology for Sustainable Development in the South (COMSATS), Pakistan;
Pontificia Universidad Cat\'{o}lica del Per\'{u}, Peru;
Ministry of Education and Science, National Science Centre and WUT ID-UB, Poland;
Korea Institute of Science and Technology Information and National Research Foundation of Korea (NRF), Republic of Korea;
Ministry of Education and Scientific Research, Institute of Atomic Physics, Ministry of Research and Innovation and Institute of Atomic Physics and Universitatea Nationala de Stiinta si Tehnologie Politehnica Bucuresti, Romania;
Ministry of Education, Science, Research and Sport of the Slovak Republic, Slovakia;
National Research Foundation of South Africa, South Africa;
Swedish Research Council (VR) and Knut \& Alice Wallenberg Foundation (KAW), Sweden;
European Organization for Nuclear Research, Switzerland;
Suranaree University of Technology (SUT), National Science and Technology Development Agency (NSTDA) and National Science, Research and Innovation Fund (NSRF via PMU-B B05F650021), Thailand;
Turkish Energy, Nuclear and Mineral Research Agency (TENMAK), Turkey;
National Academy of  Sciences of Ukraine, Ukraine;
Science and Technology Facilities Council (STFC), United Kingdom;
National Science Foundation of the United States of America (NSF) and United States Department of Energy, Office of Nuclear Physics (DOE NP), United States of America.
In addition, individual groups or members have received support from:
Czech Science Foundation (grant no. 23-07499S), Czech Republic;
European Research Council, Strong 2020 - Horizon 2020 (grant nos. 950692, 824093), European Union;
ICSC - Centro Nazionale di Ricerca in High Performance Computing, Big Data and Quantum Computing, European Union - NextGenerationEU;
Academy of Finland (Center of Excellence in Quark Matter) (grant nos. 346327, 346328), Finland.

\end{acknowledgement}

\bibliographystyle{utphys}   
\bibliography{bibliography}

\newpage
\appendix

\section{Antiparticle-to-particle ratios} \label{app:ratios}

\subsection{Particle identification of tracked species}\label{app:pid}
The particle identification (PID) of charged pions, protons, $^3\mathrm{He}$, and tritons is performed by measuring the specific energy loss ($\mathrm{d}E/\mathrm{d}x$) in both the TPC and the ITS, and particle velocity depending on the transverse momentum of the measured particles with the Time-Of-Flight detector (TOF).
Due to its electric charge $Z=2$, \Hee ~is identified using the TPC $\mathrm{d}E/\mathrm{d}x$ within $\vert n\sigma^{\mathrm{TPC}}\vert<3$, where $n\sigma$ is the deviation of the measured $\mathrm{d}E/\mathrm{d}x$ from the expected one, normalized to the experimental resolution. A similar approach is applied to protons for $p_\mathrm{T}<1\ \mathrm{GeV}/c$ after a preliminary $\vert n\sigma^{\mathrm{ITS}}\vert<3$ selection, which is required to reduce the electron contamination. Higher-$p_\mathrm{T}$ protons, pions, and tritons are identified using TOF after preliminary TPC PID selections. These PID pre-selections lead to a signal loss smaller than 0.5\% taken into account by the efficiency corrections computed via Monte Carlo simulations. The main contamination to the charged pion and proton PID is due to charged kaons and electrons. In the light-nuclei sector, the triton PID is mainly contaminated by deuterons, while tritons slightly affect the \Hee ~PID for $p_{\mathrm{T}}< 3\ \mathrm{GeV}/c$. The contribution of such misidentified tracks is estimated by fitting the corresponding $n\sigma^{\mathrm{TPC}}$ and $n\sigma^{\mathrm{TOF}}$ distributions. The $n\sigma^{\mathrm{TPC}}$ or $n\sigma^{\mathrm{TOF}}$ background distributions are fitted outside of the signal window. Their extrapolation within the signal region is integrated to statistically subtract the contamination due to either misidentified particles or the mismatch of TPC tracks and TOF space points.

\subsection{Machine Learning analysis of \texorpdfstring{$\Omega$}{Omega} and \texorpdfstring{$^{3}_{\Lambda}\mathrm{H}$}{H}}
\label{app:ml}
The Machine Learning (ML) selection of $\Omega$ and $^{3}_{\Lambda}\mathrm{H}$ candidates is based on Boosted Decision Trees (BDT). The optimization of the BDT internal parameters is performed using samples of correctly classified signal and background candidates.
The signal sample is built from simulated candidates injected on top of a HIJING Pb--Pb event with a Blast-wave $p_\mathrm{T}$ distribution~\cite{Schnedermann:1993ws} derived from the measured production of light flavor hadrons for $\Omega$~\cite{ALICE:2019hno} and of $^3\mathrm{He}$ for $^{3}_{\Lambda}\mathrm{H}$~\cite{ALICE:2022veq}. The background $^{3}_{\Lambda}\mathrm{H}$ candidates are obtained in the data from same-sign combinations of $^3\mathrm{He}$ and $\pi$ tracks. For $\Omega^-$, all the candidates with an invariant-mass deviating more than $7\sigma$ from the nominal $\Omega^-$ mass are considered as background candidates, with $\sigma\approx 1.7\ \mathrm{MeV}/c^2$ being the invariant-mass resolution in the data. The same training variables used in previous analyses are employed for ${}^3_{\Lambda}\mathrm{H}$~\cite{ALargeIonColliderExperiment:2021puh, ALICE:2022sco}: the cosine of the pointing angle $\cos(\theta_p)$ (i.e., the angle between the reconstructed candidate momentum and the straight line connecting the production and decay vertices), the DCA between the decay tracks and the primary vertex (PV), and between the two tracks themselves, the number of TPC space points for the ${}^{3}\mathrm{He}$ track, and the $n\sigma^{\mathrm{TPC}}$ of the decay tracks. For the $\Omega^-$, the BDT input variables include the DCA of the $\mathrm{K^{-}}$, $\pi^{-}$ and $\mathrm{p}$ to the PV, the DCA of the reconstructed $\Lambda$ to the PV, the minimum distance between the $\pi^{-}$ and $\mathrm{p}$, and between the $\mathrm{K^-}$ and $\Lambda$. The $\cos(\theta_p)$ for both the $\Omega^-$ and $\Lambda$, and the $n\sigma^{\mathrm{TPC}}$ for $\mathrm{p}$, are also used as BDT input variables. 
Signal candidates are selected requiring a BDT output score larger than a preset threshold. For $^3_{\Lambda}\mathrm{H}$, the threshold is optimized by maximizing the expected signal significance; for the $\Omega^{-}$, a BDT signal selection efficiency of 50\% is required, as it ensures a consistent BDT response in data and MC. 

\subsection{Efficiency and absorption corrections}

The efficiency corrections applied in this Letter take into account the tracking and candidate-selection efficiencies, including PID. Charged pions, protons, and tritons are selected with an efficiency of about 40\% when requiring TOF PID, while protons having $p_{\mathrm{T}}<1\ \mathrm{GeV}/c$ and \Hee~ candidates are selected with an efficiency of approximately 70\%. For antiproton, antitriton, and $^3\overline{\mathrm{He}}$ candidates, the obtained efficiencies are lower than those of their respective charge-conjugates by about 5\% to 10\% due to their larger absorption cross sections inside the ALICE apparatus. For $\Omega^{-}$ baryons, the efficiency of the preliminary selections is about 5\%, while the BDT signal-selection efficiency is about 50\%. For $^3_{\Lambda}\mathrm{H}$, these two efficiencies are about 30\% and 70\%, respectively.

The modeling of the absorption cross sections in GEANT 4 is improved by using the available measurements of absorption cross sections in different materials~\cite{Chen1955CH67, Booth1957_exforO2097.1, Porile1962PO12, Artun1975AR17, Sadler1980SA13, Ashery:1981tq, Nakamura:1984xw, McGill1984MC02, Zaitseva1992ZA02, Kuzichev:1993tn, Herbach2006HE16, Zamani2010ZA05, Bardayan2010BA21, ALICE:2022zuz, ALICE:2023aod}. Dedicated correction factors are computed as the ratios between the efficiencies computed using either the default parameterizations or those re-tuned on the experimental data. For charged pions and protons, the measurements obtained in lower energy experiments are used, while for tritons and \Hee~ the available ALICE measurements are employed. For the light (anti)nuclei, the observed effect is between 1\% and 3\% across the analysed $p_\mathrm{T}$ range. The uncertainties on the measured cross sections are propagated to the antiparticle-to-particle ratios as centrality-correlated sources of systematic uncertainties. The resulting uncertainties are reported in Table~\ref{tab:sys}. For the $^3_{\Lambda}\mathrm{H}$, an absorption factor is computed to correct for undetected hypernuclei that are absorbed in the apparatus before decaying. A sample of simulated \Hee~ candidates is used to approximate the absorption of $^3_{\Lambda}\mathrm{H}$ inside the ALICE apparatus. The absorbed fraction is about 6\% and 4\% for $^3_{\overline{\Lambda}}\mathrm{\overline{H}}$ and $^3_{\Lambda}\mathrm{H}$, respectively.

\subsection{Systematic uncertainties}\label{app:sys}
The centrality-uncorrelated systematic uncertainty on the yield ratio is obtained as the variance of multiple reanalyses done by varying the tracking and PID selections for $\pi$, $\mathrm{p}$, $^3\mathrm{He}$, and $^3\mathrm{H}$, and of the BDT output selections for $^3_{\Lambda}\mathrm{H}$ and $\Omega^-$ around their nominal values used in the analysis. The background fit function is also changed from exponential to polynomial in the invariant mass fit of $^3_{\Lambda}\mathrm{H}$ and $\Omega^-$, while the yields of $^3\mathrm{He}$ and  $^3\mathrm{H}$ are alternatively extracted as the integral of a Gaussian fit to the $n\sigma^{\mathrm{TPC}}$ and $n\sigma^{\mathrm{TOF}}$ distributions, respectively. The variations are applied coherently to antiparticles and particles to allow for the cancelation of correlated contributions in the antiparticle-to-particle yield ratios. The MC statistical precision is also considered as a centrality-uncorrelated source of systematic uncertainty. The uncertainty on the material-budget description in MC simulations is correlated with centrality. It is evaluated by varying the amount of material crossed by simulated particles by its uncertainty, estimated to be $\pm4.5\%$~\cite{ALICE:2014sbx}. The uncertainties on the measured absorption cross sections used to correct the GEANT4 ones are also propagated to the ratios. The consistency of the results obtained with opposite magnetic field polarities is assessed by repeating the measurement separately with the two configurations: a statistically significant discrepancy of about $0.4\%$ and $0.6\%$ due to imperfections in the MC description is observed in 
semicentral and central collisions,
respectively. The maximum half dispersion between the opposite field polarity results is then assigned as a further centrality-correlated uncertainty. The values of the various contributions are summarised in Table~\ref{tab:sys}.

\begin{table}[!htbp]
\begin{center}
\caption{Relative systematic uncertainty on the average antiparticle-to-particle ratios due to the different sources considered in the analysis. Only the statistically significant contributions to systematic uncertainties are reported in the table.}
\begin{tabular}{ l c c c c c c } \hline
Source & $\overline{\Omega}^{+}/\Omega^{-}$ & $\pi^+/\pi^-$ & $\overline{\mathrm{p}}/\mathrm{p}$ & ${}^{3}_{\overline{\Lambda}}\overline{\mathrm{H}}/{}^{3}_{\Lambda}\mathrm{H}$ & ${}^{3}\overline{\mathrm{H}}/{}^{3}\mathrm{H}$ & ${}^{3}\overline{\mathrm{He}}/{}^{3}\mathrm{He}$ \\ \hline
Candidate selection + signal extract. & 0.5\% & 0.05\% & 0.05\% & 10\% & 3\% & 0.5\% \\
Monte Carlo precision & 0.5\% & 0.1\% & 0.1\% & 1\% & 1\% & 1\% \\
Material budget & -- & 0.1\% & 0.5\% & -- & -- & -- \\
Absorption cross section & -- & 0.7\% & 0.5\% & 1\% & 10\% & 1\% \\
Magnetic field polarity & -- & 0.2--0.3\% & 0.2--0.3\% & -- & -- & -- \\ \hline
\end{tabular}
\label{tab:sys}
\end{center}
\end{table}

\subsection{Results for the antiparticle-to-particle ratios}\label{app:ratiosResults}
\begin{figure}[!htbp]
    \centering
    \includegraphics[width=0.32\textwidth]{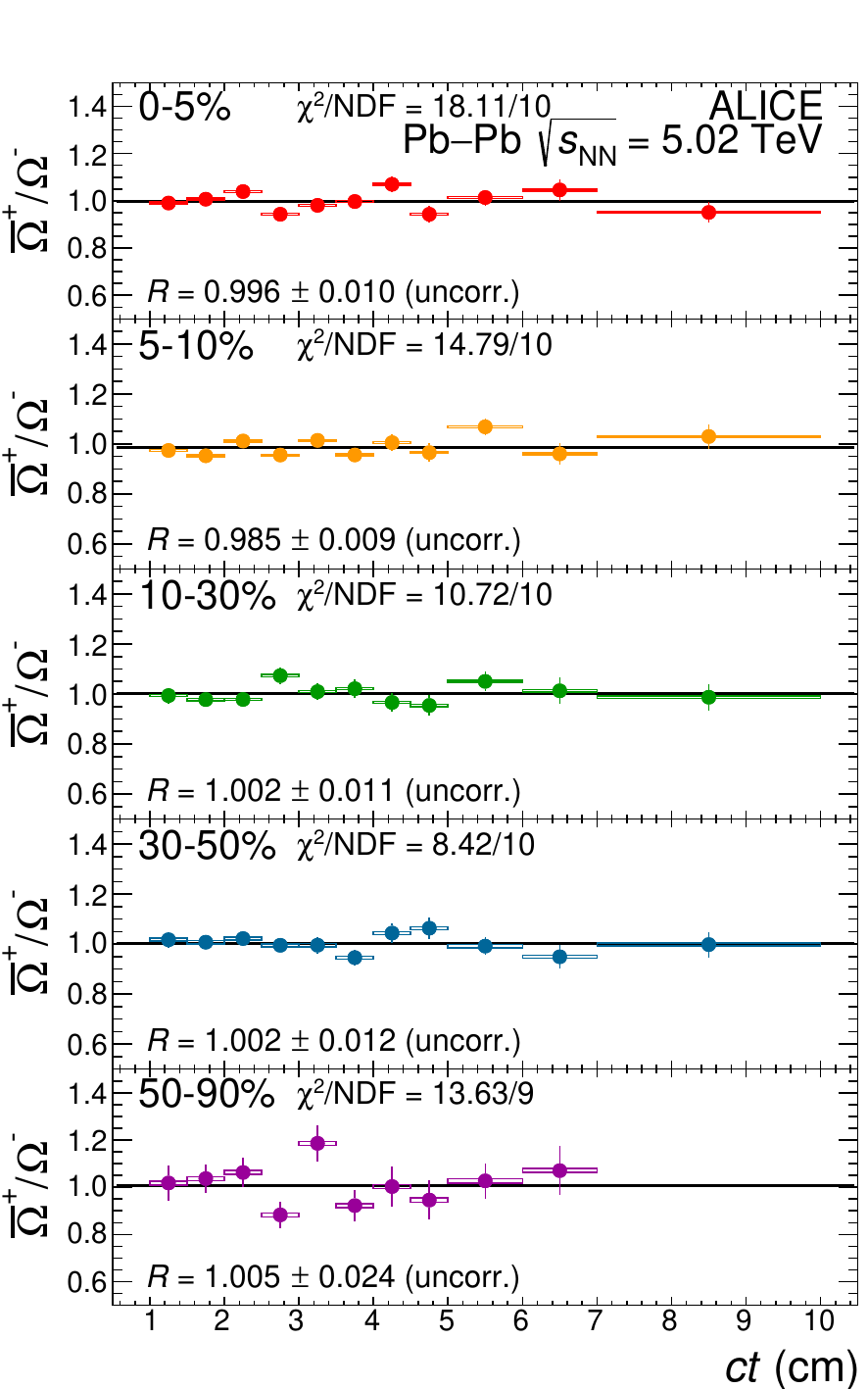}
    \includegraphics[width=0.32\textwidth]{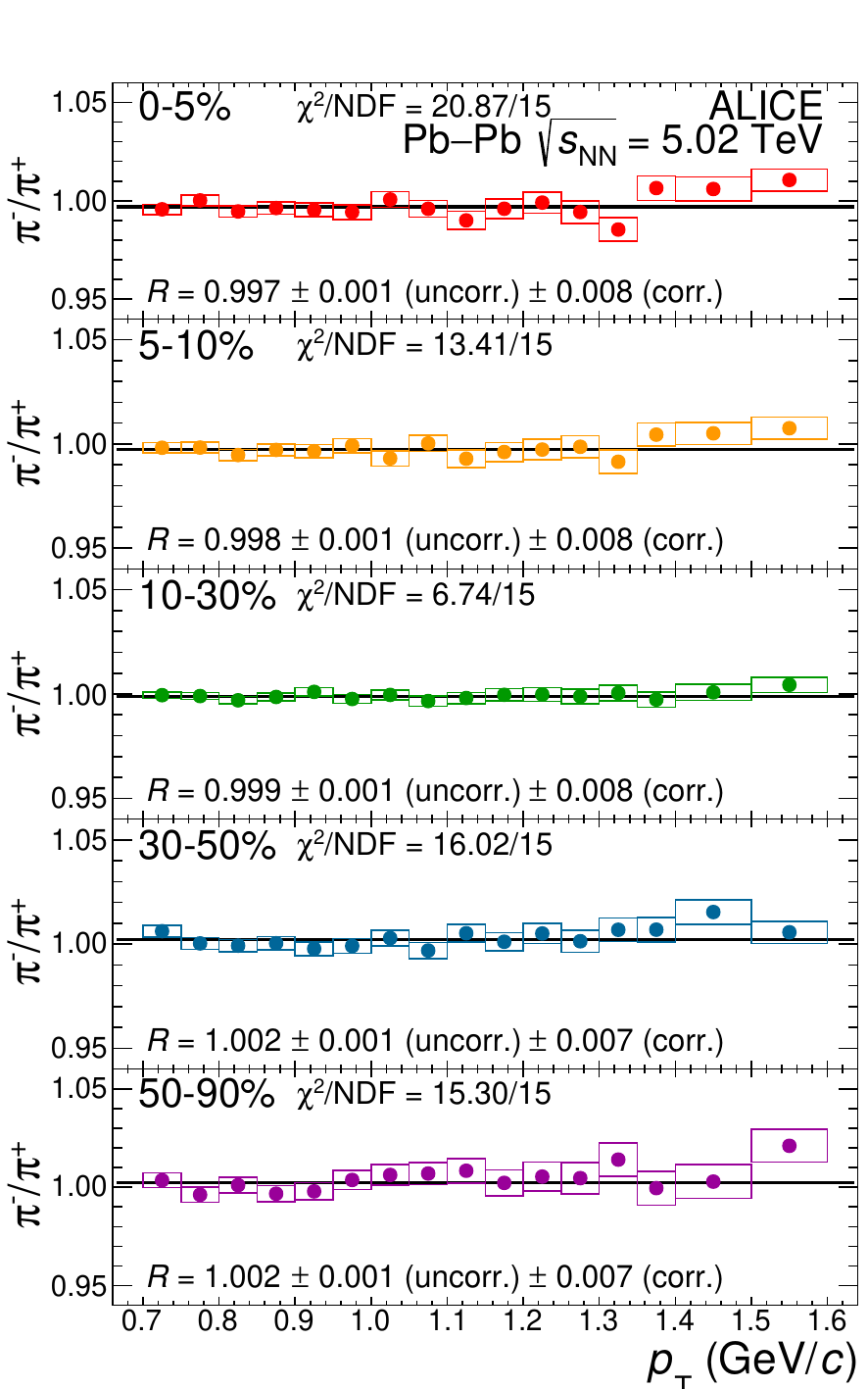}
    \includegraphics[width=0.32\textwidth]{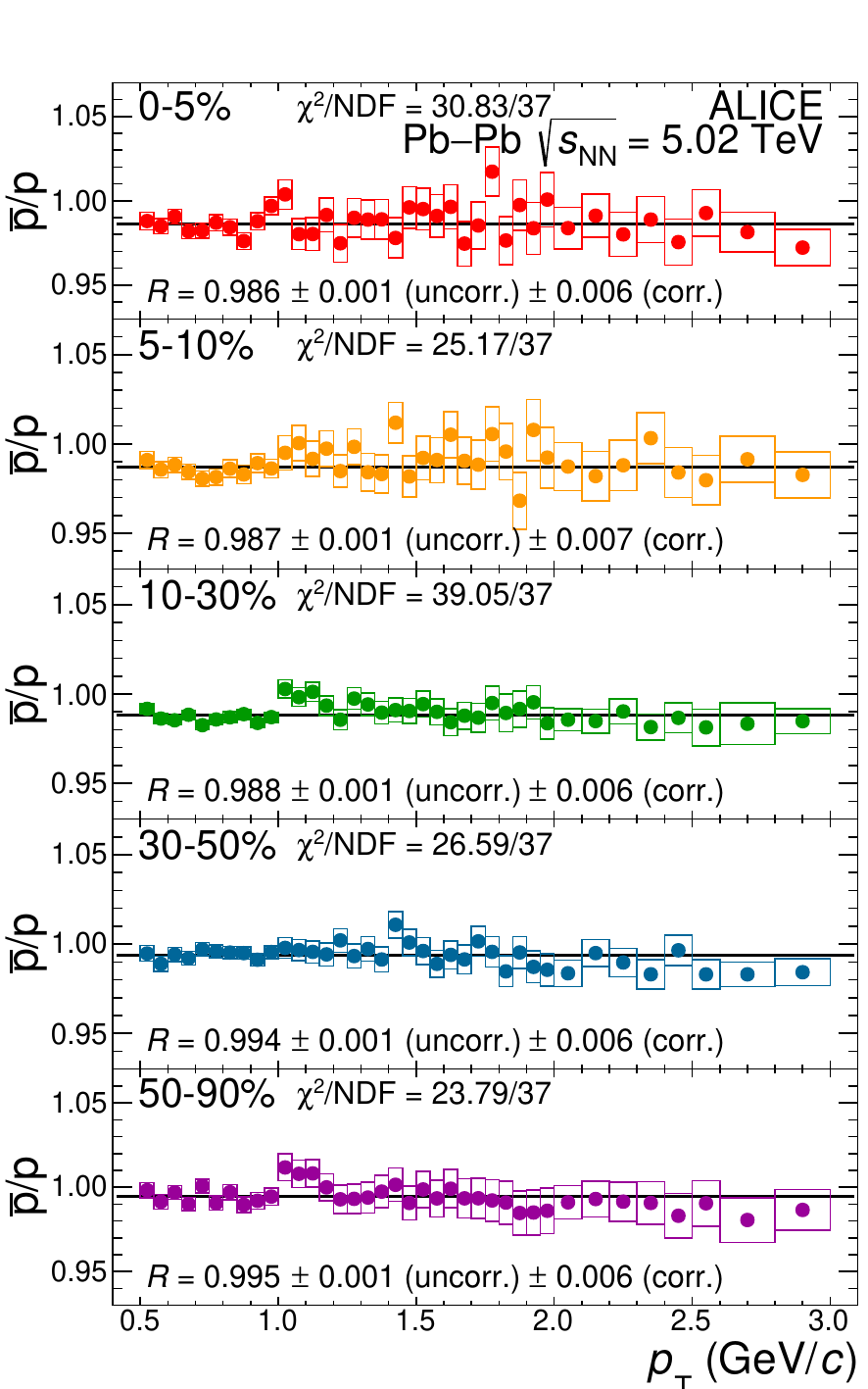}
    \includegraphics[width=0.32\textwidth]{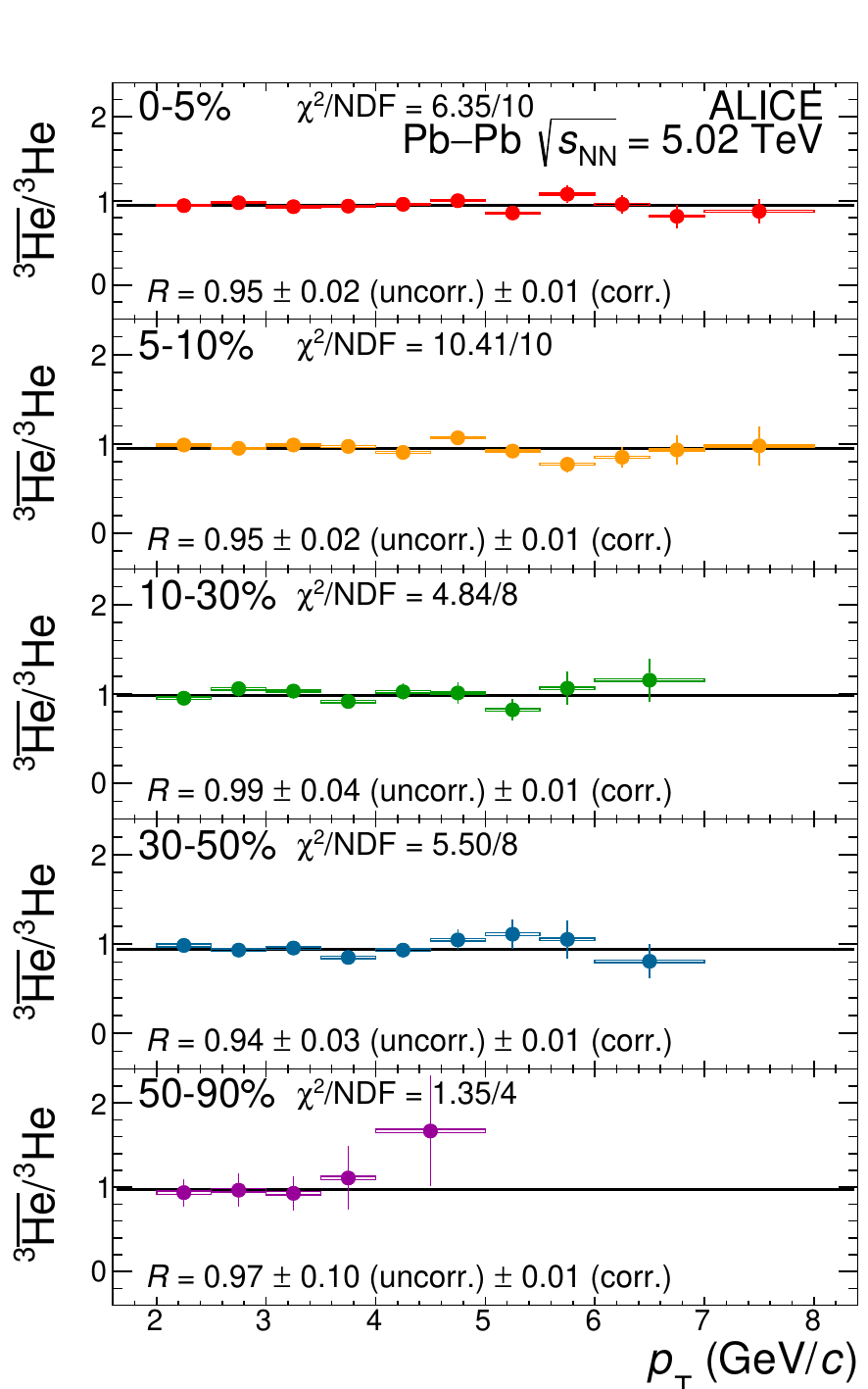}
    \includegraphics[width=0.32\textwidth]{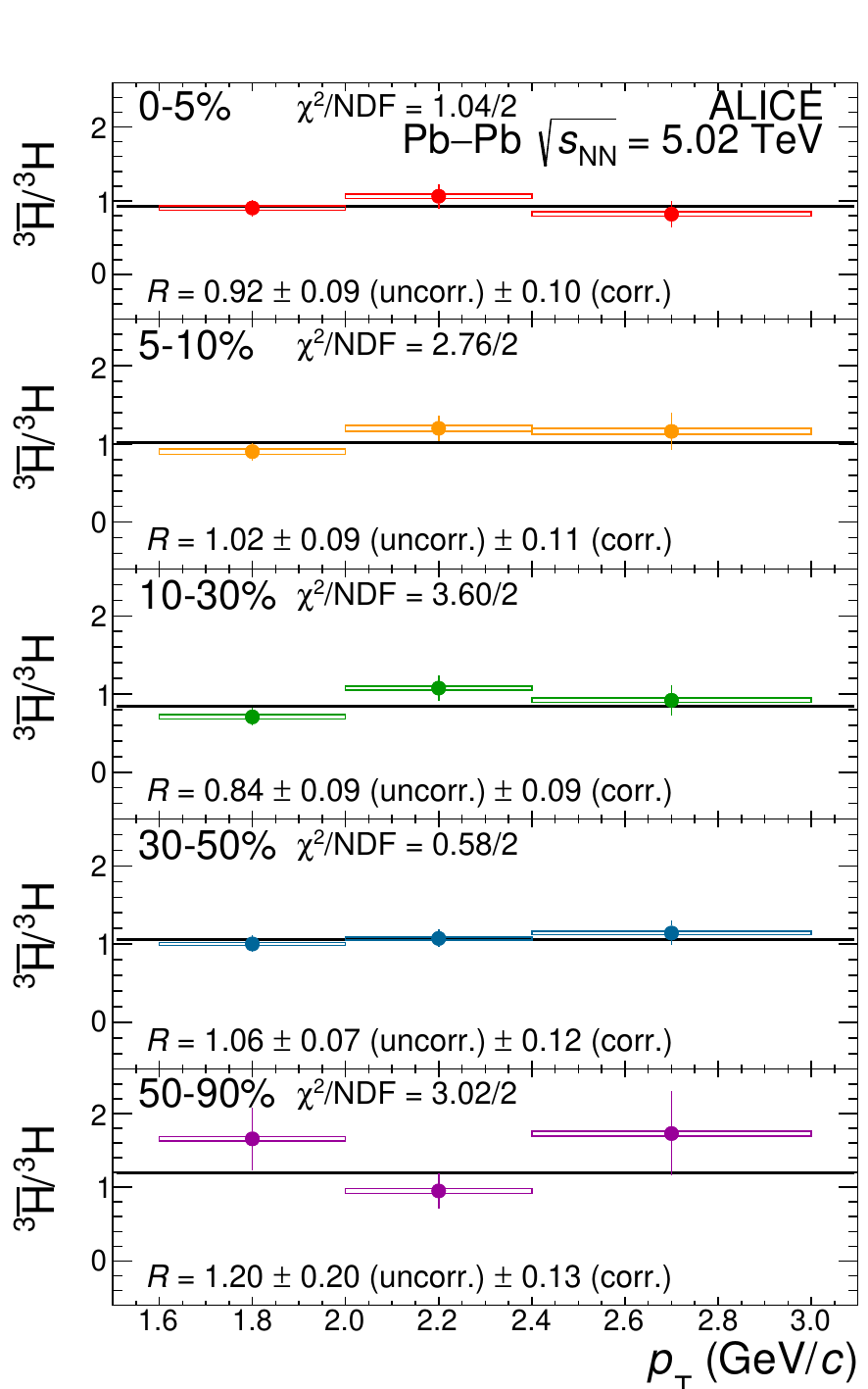}
    \includegraphics[width=0.32\textwidth]{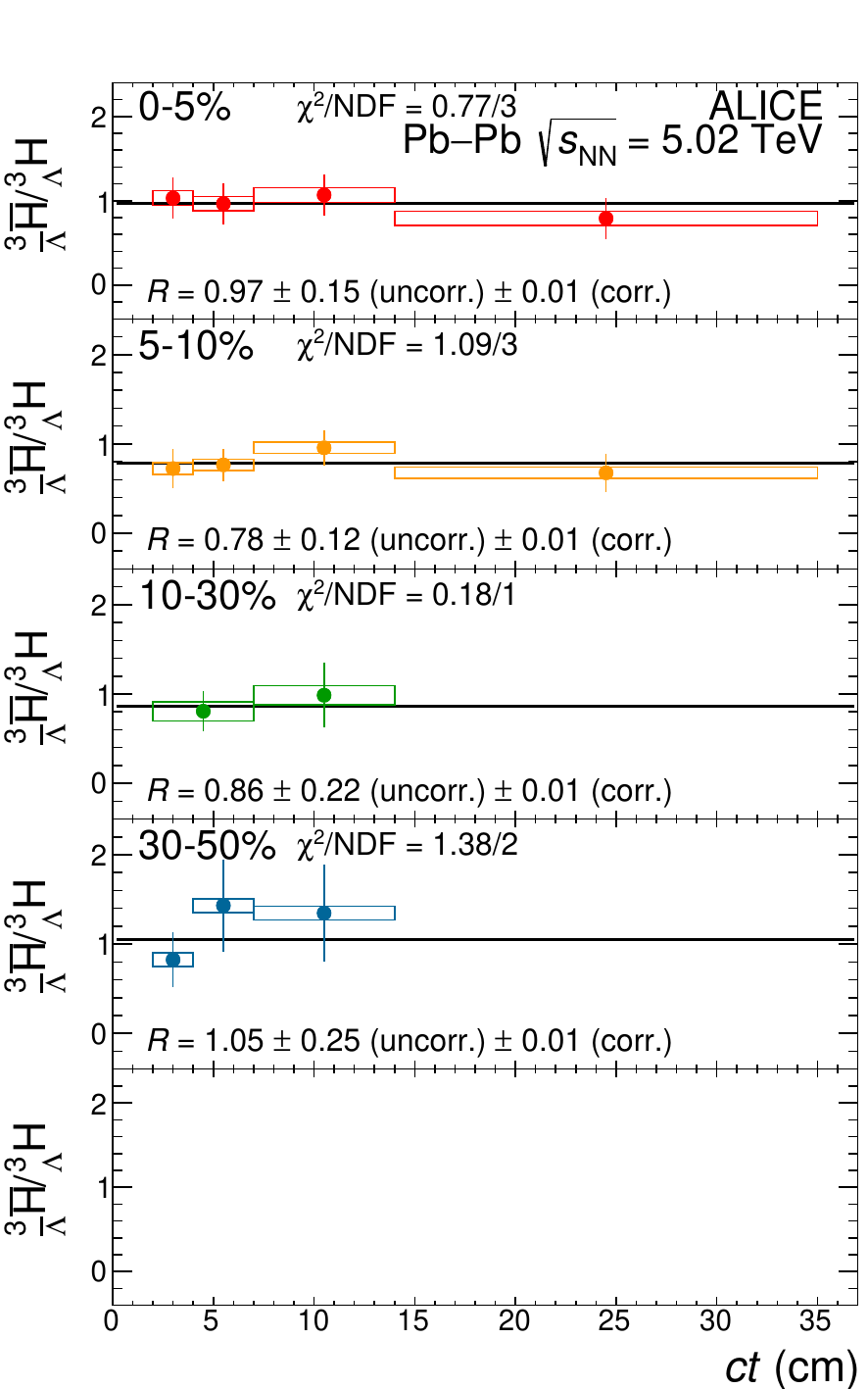}
    \caption{$p_{\mathrm{T}}$- and $ct$-differential ratios of the species used for the chemical potential measurement in the various centrality intervals. Error bars show statistical uncertainties, while boxes represent centrality-uncorrelated uncertainties. The value of $R$ represents the averages weighted with the total uncorrelated uncertainties of the differential measurements. The correlated uncertainties are not shown in the plots.
}
    \label{fig:ratios}
\end{figure}

\section{Covariance matrices} \label{app:covariances}
\begin{figure}[h]
    \centering
    \includegraphics[width=0.49\textwidth]{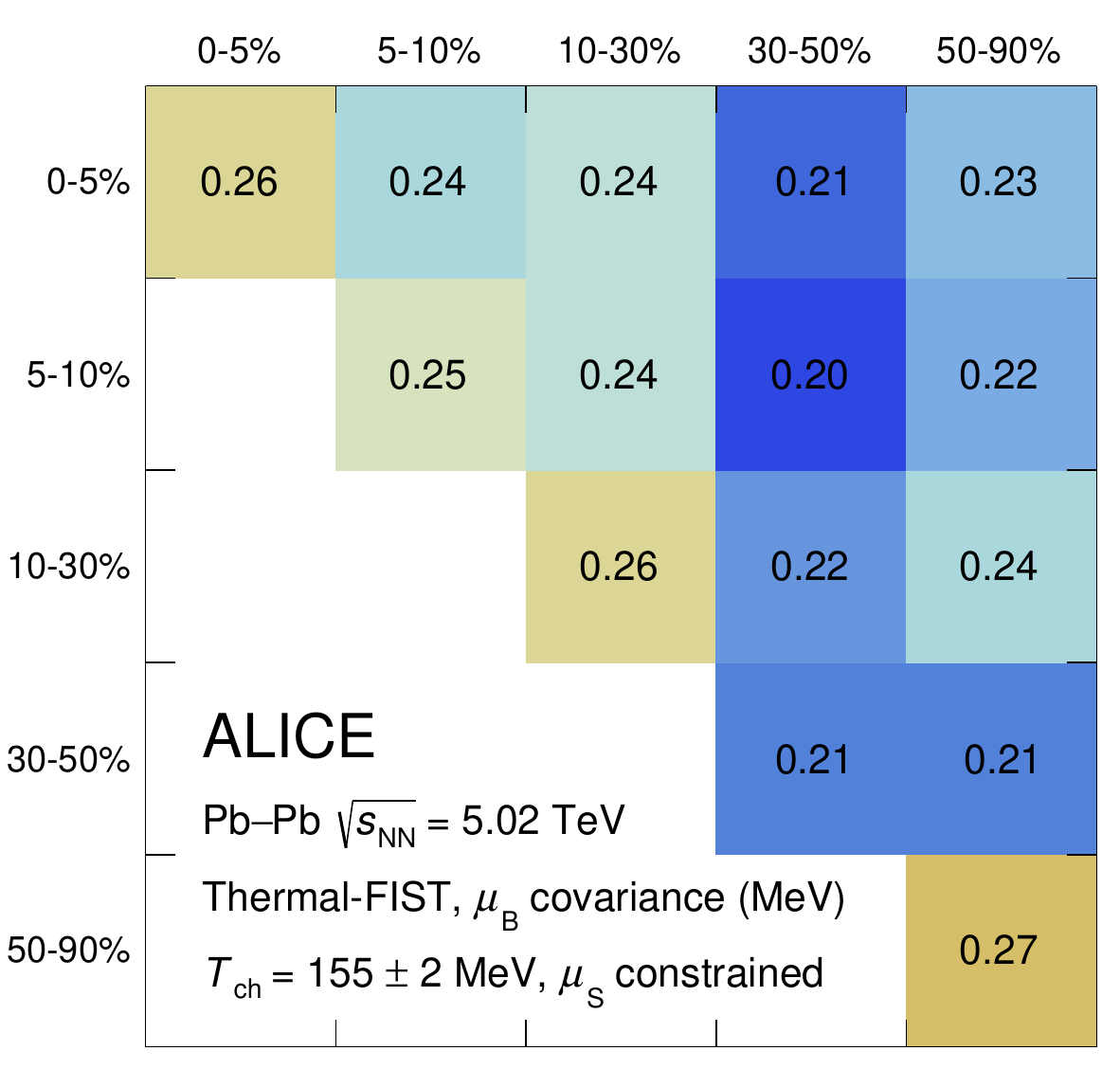}
    \includegraphics[width=0.49\textwidth]{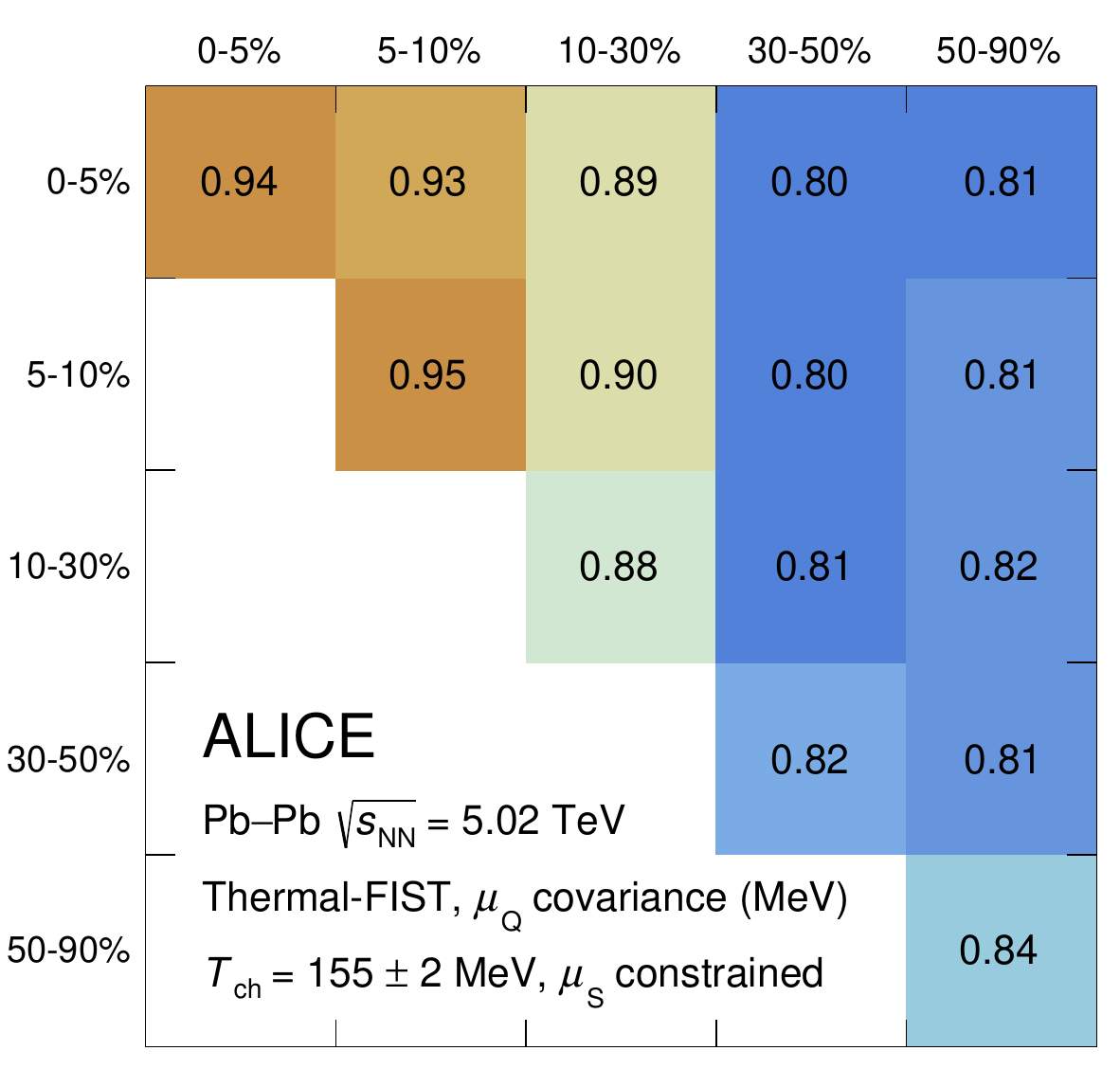}
    \caption{Covariance matrices of $\mu_\mathrm{B}$ (left) and $\mu_\mathrm{Q}$ (right) obtained with Thermal-FIST~\cite{Vovchenko:2019pjl} in the different centrality intervals.}
    \label{fig:covariances}
\end{figure}
\section{\texorpdfstring{$\chi^{2}$}{Chi2} profiles of chemical potential fits} \label{app:profiles}

\subsection{Thermal-FIST model}\label{app:chi2Thefist}

\begin{table}[!htbp]
\begin{center}
\caption{Confidence intervals (C.I.) at $1\sigma$, $2\sigma$, and $3\sigma$ levels for $\mu_B$ and $\mu_Q$.}
\begin{tabular}{ l c c c } \hline
Observable & $1\sigma$ C.I. (MeV) & $2\sigma$ C.I. (MeV) & $3\sigma$ C.I. (MeV) \\ \hline
$\mu_B$ & $[0.26, 1.16]$ & $[-0.20, 1.61]$ & $[-0.65, 2.07]$ \\
$\mu_Q$ & $[-1.08, 0.73]$ & $[-1.98, 1.63]$ & $[-2.89, 2.54]$ \\ \hline
\end{tabular}
\end{center}
\end{table}

\begin{figure}[!htbp]
    \centering
    \includegraphics[width=0.49\textwidth]{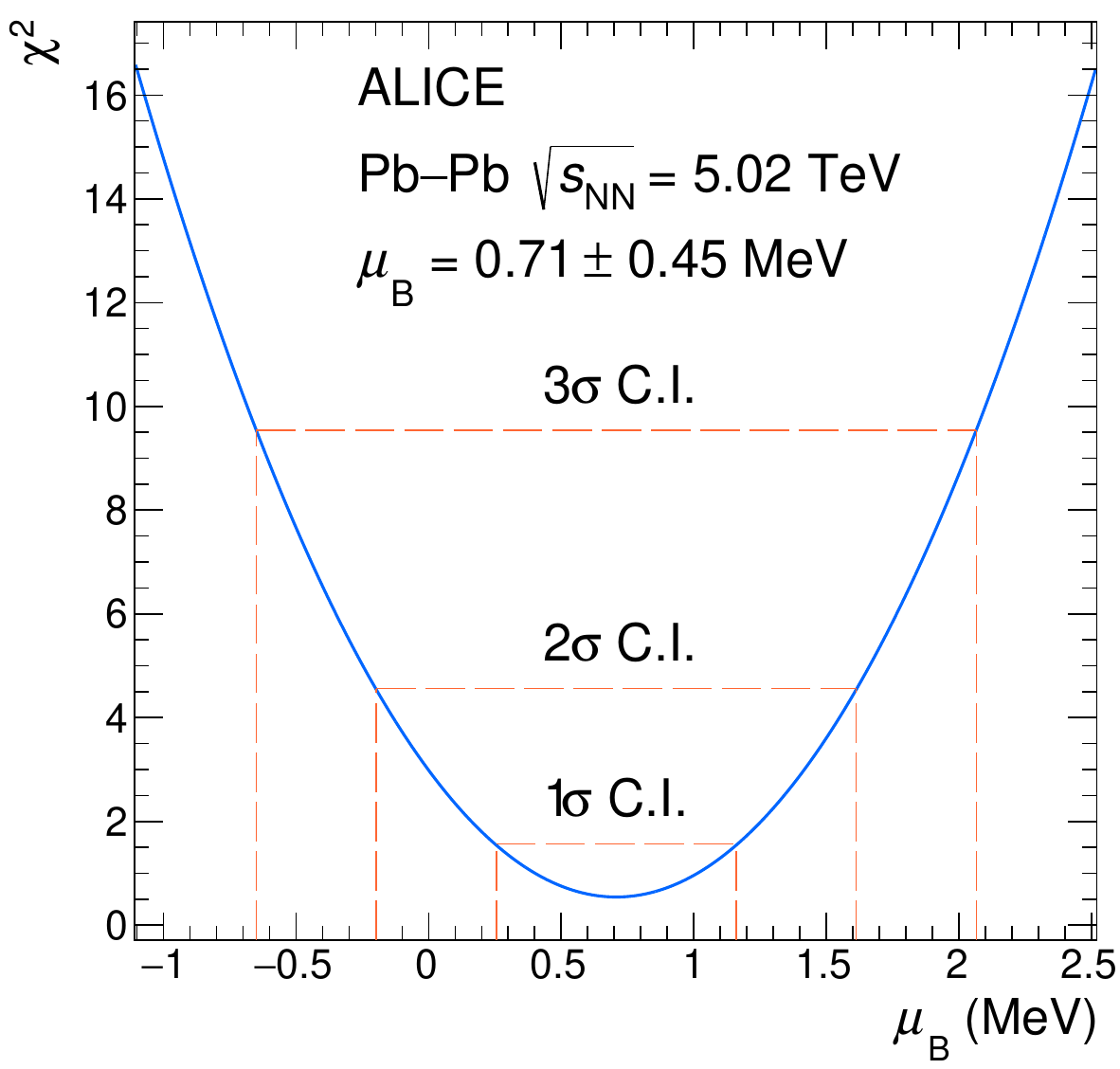}
    \includegraphics[width=0.49\textwidth]{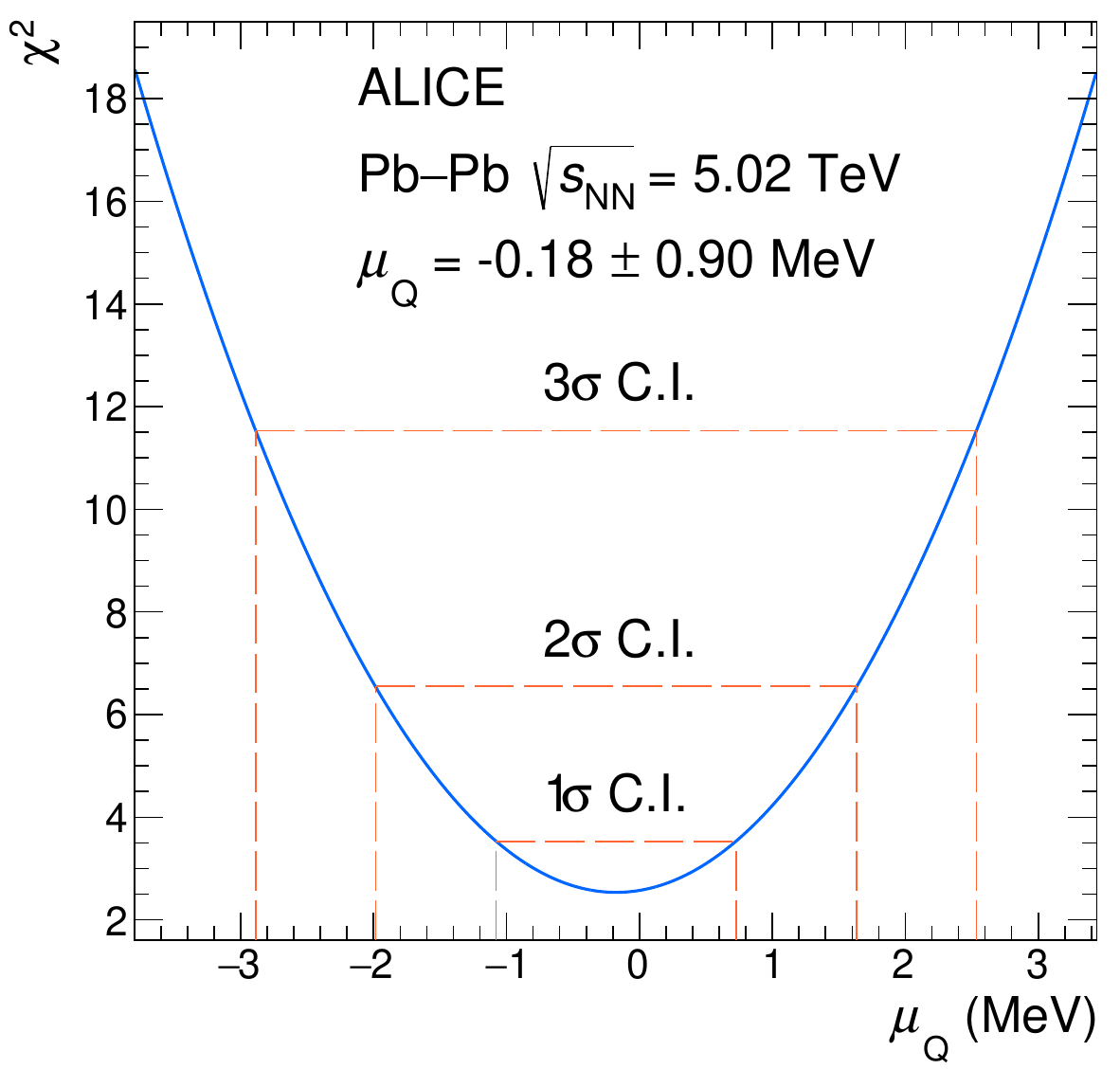}
    \caption{Profiles of the $\chi^{2}$ variable minimized in the fit of $\mu_B$ (left) and $\mu_Q$ (right) obtained with the Thermal-FIST model~\cite{Vovchenko:2019pjl}. The values obtained from the minimization, as well as the $1\sigma$, $2\sigma$, and $3\sigma$ confidence intervals, are reported in the figures.}
    \label{fig:profiles}
\end{figure}

\subsection{GSI-Heidelberg model}\label{app:chi2GSI}

\begin{table}[!htbp]
\begin{center}
\caption{Confidence intervals (C.I.) at $1\sigma$, $2\sigma$, and $3\sigma$ levels for $\mu_B$ and $\mu_Q$.}
\begin{tabular}{ l c c c } \hline
Observable & $1\sigma$ C.I. (MeV) & $2\sigma$ C.I. (MeV) & $3\sigma$ C.I. (MeV) \\ \hline
$\mu_B$ & $[0.47, 1.33]$ & $[0.05, 1.75]$ & $[-0.38, 2.18]$ \\ \hline
\end{tabular}
\end{center}
\end{table}

\begin{figure}[!htbp]
    \centering
    \includegraphics[width=0.49\textwidth]{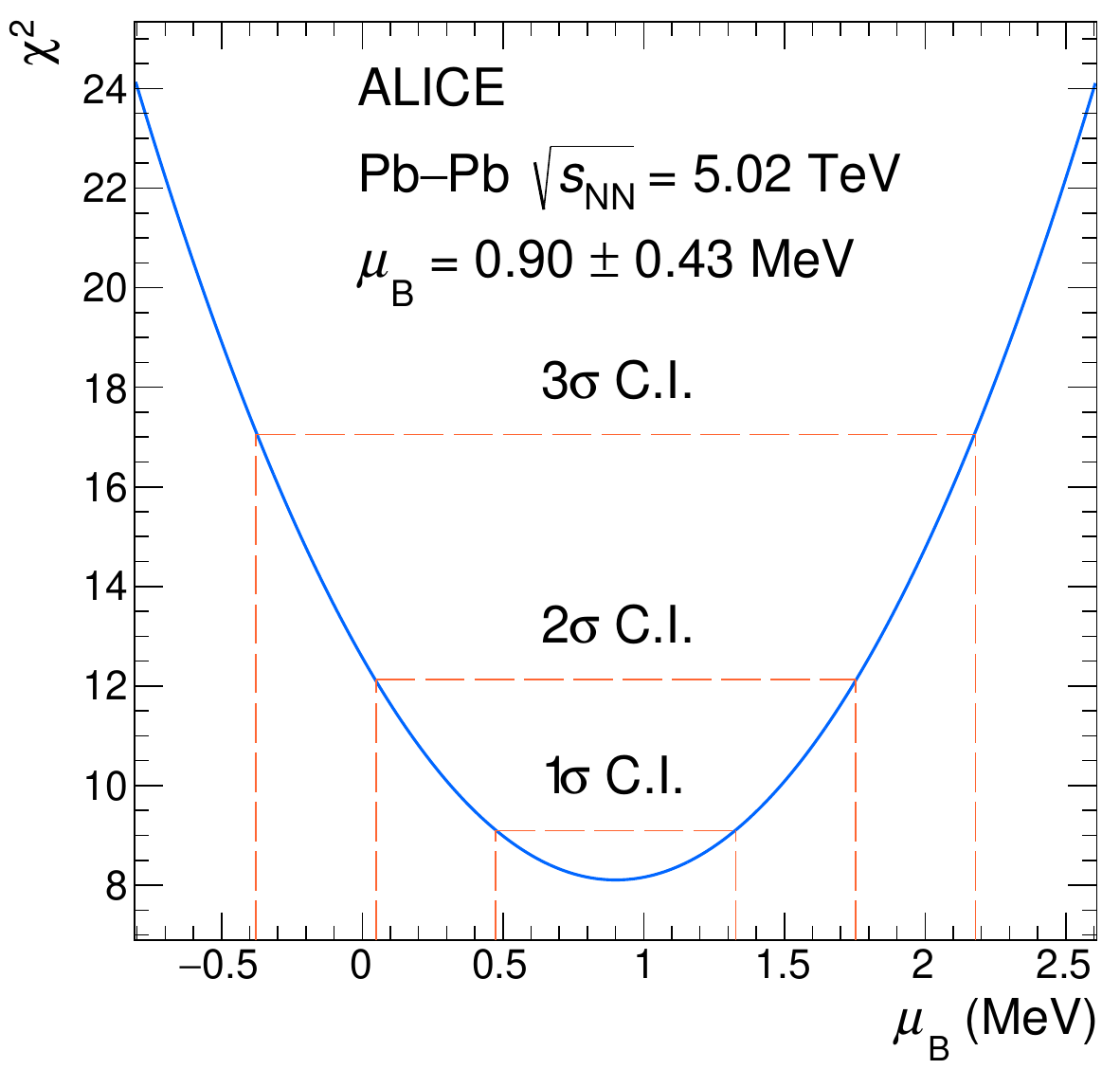}
    \caption{Profiles of the $\chi^{2}$ variable minimized in the fit of $\mu_B$ obtained with the GSI-Heidelberg model~\cite{Braun-Munzinger:2003pwq, Andronic:2005yp, Andronic:2018qqt}. The values obtained from the minimization, as well as the $1\sigma$, $2\sigma$, and $3\sigma$ confidence intervals, are reported in the figures.}
    \label{fig:profilesGSI}
\end{figure}

%
%

\section{The ALICE Collaboration}
\label{app:collab}
\begin{flushleft} 
\small

S.~Acharya\,\orcidlink{0000-0002-9213-5329}\,$^{\rm 128}$, 
D.~Adamov\'{a}\,\orcidlink{0000-0002-0504-7428}\,$^{\rm 87}$, 
G.~Aglieri Rinella\,\orcidlink{0000-0002-9611-3696}\,$^{\rm 33}$, 
L.~Aglietta$^{\rm 25}$, 
M.~Agnello\,\orcidlink{0000-0002-0760-5075}\,$^{\rm 30}$, 
N.~Agrawal\,\orcidlink{0000-0003-0348-9836}\,$^{\rm 26}$, 
Z.~Ahammed\,\orcidlink{0000-0001-5241-7412}\,$^{\rm 136}$, 
S.~Ahmad\,\orcidlink{0000-0003-0497-5705}\,$^{\rm 16}$, 
S.U.~Ahn\,\orcidlink{0000-0001-8847-489X}\,$^{\rm 72}$, 
I.~Ahuja\,\orcidlink{0000-0002-4417-1392}\,$^{\rm 38}$
% ...

\section*{Affiliation Notes}

$^{\rm I}$ Deceased\\
$^{\rm II}$ Also at: Max-Planck-Institut fur Physik, Munich, Germany\\
$^{\rm III}$ Also at: Italian National Agency for New Technologies, Energy and Sustainable Economic Development (ENEA), Bologna, Italy\\
$^{\rm IV}$ Also at: Dipartimento DET del Politecnico di Torino, Turin, Italy\\
$^{\rm V}$ Also at: Yildiz Technical University, Istanbul, T\"{u}rkiye\\
$^{\rm VI}$ Also at: Department of Applied Physics, Aligarh Muslim University, Aligarh, India\\
$^{\rm VII}$ Also at: Institute of Theoretical Physics, University of Wroclaw, Poland\\
$^{\rm VIII}$ Also at: An institution covered by a cooperation agreement with CERN\\

\section*{Collaboration Institutes}

$^{1}$ A.I. Alikhanyan National Science Laboratory (Yerevan Physics Institute) Foundation, Yerevan, Armenia\\
$^{2}$ AGH University of Krakow, Cracow, Poland\\
$^{3}$ Bogolyubov Institute for Theoretical Physics, National Academy of Sciences of Ukraine, Kiev, Ukraine\\
$^{4}$ Bose Institute, Department of Physics  and Centre for Astroparticle Physics and Space Science (CAPSS), Kolkata, India\\
$^{5}$ California Polytechnic State University, San Luis Obispo, California, United States\\
$^{6}$ Central China Normal University, Wuhan, China\\
$^{7}$ Centro de Aplicaciones Tecnol\'{o}gicas y Desarrollo Nuclear (CEADEN), Havana, Cuba\\
$^{8}$ Centro de Investigaci\'{o}n y de Estudios Avanzados (CINVESTAV), Mexico City and M\'{e}rida, Mexico\\
$^{9}$ Chicago State University, Chicago, Illinois, United States\\
$^{10}$ China Institute of Atomic Energy, Beijing, China\\
$^{11}$ China University of Geosciences, Wuhan, China\\
$^{12}$ Chungbuk National University, Cheongju, Republic of Korea\\
$^{13}$ Comenius University Bratislava, Faculty of Mathematics, Physics and Informatics, Bratislava, Slovak Republic\\
$^{14}$ COMSATS University Islamabad, Islamabad, Pakistan\\
$^{15}$ Creighton University, Omaha, Nebraska, United States\\
$^{16}$ Department of Physics, Aligarh Muslim University, Aligarh, India\\
$^{17}$ Department of Physics, Pusan National University, Pusan, Republic of Korea\\
$^{18}$ Department of Physics, Sejong University, Seoul, Republic of Korea\\
$^{19}$ Department of Physics, University of California, Berkeley, California, United States\\
$^{20}$ Department of Physics, University of Oslo, Oslo, Norway\\
$^{21}$ Department of Physics and Technology, University of Bergen, Bergen, Norway\\
$^{22}$ Dipartimento di Fisica, Universit\`{a} di Pavia, Pavia, Italy\\
$^{23}$ Dipartimento di Fisica dell'Universit\`{a} and Sezione INFN, Cagliari, Italy\\
$^{24}$ Dipartimento di Fisica dell'Universit\`{a} and Sezione INFN, Trieste, Italy\\
$^{25}$ Dipartimento di Fisica dell'Universit\`{a} and Sezione INFN, Turin, Italy\\
$^{26}$ Dipartimento di Fisica e Astronomia dell'Universit\`{a} and Sezione INFN, Bologna, Italy\\
$^{27}$ Dipartimento di Fisica e Astronomia dell'Universit\`{a} and Sezione INFN, Catania, Italy\\
$^{28}$ Dipartimento di Fisica e Astronomia dell'Universit\`{a} and Sezione INFN, Padova, Italy\\
$^{29}$ Dipartimento di Fisica `E.R.~Caianiello' dell'Universit\`{a} and Gruppo Collegato INFN, Salerno, Italy\\
$^{30}$ Dipartimento DISAT del Politecnico and Sezione INFN, Turin, Italy\\
$^{31}$ Dipartimento di Scienze MIFT, Universit\`{a} di Messina, Messina, Italy\\
$^{32}$ Dipartimento Interateneo di Fisica `M.~Merlin' and Sezione INFN, Bari, Italy\\
$^{33}$ European Organization for Nuclear Research (CERN), Geneva, Switzerland\\
$^{34}$ Faculty of Electrical Engineering, Mechanical Engineering and Naval Architecture, University of Split, Split, Croatia\\
$^{35}$ Faculty of Engineering and Science, Western Norway University of Applied Sciences, Bergen, Norway\\
$^{36}$ Faculty of Nuclear Sciences and Physical Engineering, Czech Technical University in Prague, Prague, Czech Republic\\
$^{37}$ Faculty of Physics, Sofia University, Sofia, Bulgaria\\
$^{38}$ Faculty of Science, P.J.~\v{S}af\'{a}rik University, Ko\v{s}ice, Slovak Republic\\
$^{39}$ Frankfurt Institute for Advanced Studies, Johann Wolfgang Goethe-Universit\"{a}t Frankfurt, Frankfurt, Germany\\
$^{40}$ Fudan University, Shanghai, China\\
$^{41}$ Gangneung-Wonju National University, Gangneung, Republic of Korea\\
$^{42}$ Gauhati University, Department of Physics, Guwahati, India\\
$^{43}$ Helmholtz-Institut f\"{u}r Strahlen- und Kernphysik, Rheinische Friedrich-Wilhelms-Universit\"{a}t Bonn, Bonn, Germany\\
$^{44}$ Helsinki Institute of Physics (HIP), Helsinki, Finland\\
$^{45}$ High Energy Physics Group,  Universidad Aut\'{o}noma de Puebla, Puebla, Mexico\\
$^{46}$ Horia Hulubei National Institute of Physics and Nuclear Engineering, Bucharest, Romania\\
$^{47}$ HUN-REN Wigner Research Centre for Physics, Budapest, Hungary\\
$^{48}$ Indian Institute of Technology Bombay (IIT), Mumbai, India\\
$^{49}$ Indian Institute of Technology Indore, Indore, India\\
$^{50}$ INFN, Laboratori Nazionali di Frascati, Frascati, Italy\\
$^{51}$ INFN, Sezione di Bari, Bari, Italy\\
$^{52}$ INFN, Sezione di Bologna, Bologna, Italy\\
$^{53}$ INFN, Sezione di Cagliari, Cagliari, Italy\\
$^{54}$ INFN, Sezione di Catania, Catania, Italy\\
$^{55}$ INFN, Sezione di Padova, Padova, Italy\\
$^{56}$ INFN, Sezione di Pavia, Pavia, Italy\\
$^{57}$ INFN, Sezione di Torino, Turin, Italy\\
$^{58}$ INFN, Sezione di Trieste, Trieste, Italy\\
$^{59}$ Inha University, Incheon, Republic of Korea\\
$^{60}$ Institute for Gravitational and Subatomic Physics (GRASP), Utrecht University/Nikhef, Utrecht, Netherlands\\
$^{61}$ Institute of Experimental Physics, Slovak Academy of Sciences, Ko\v{s}ice, Slovak Republic\\
$^{62}$ Institute of Physics, Homi Bhabha National Institute, Bhubaneswar, India\\
$^{63}$ Institute of Physics of the Czech Academy of Sciences, Prague, Czech Republic\\
$^{64}$ Institute of Space Science (ISS), Bucharest, Romania\\
$^{65}$ Institut f\"{u}r Kernphysik, Johann Wolfgang Goethe-Universit\"{a}t Frankfurt, Frankfurt, Germany\\
$^{66}$ Instituto de Ciencias Nucleares, Universidad Nacional Aut\'{o}noma de M\'{e}xico, Mexico City, Mexico\\
$^{67}$ Instituto de F\'{i}sica, Universidade Federal do Rio Grande do Sul (UFRGS), Porto Alegre, Brazil\\
$^{68}$ Instituto de F\'{\i}sica, Universidad Nacional Aut\'{o}noma de M\'{e}xico, Mexico City, Mexico\\
$^{69}$ iThemba LABS, National Research Foundation, Somerset West, South Africa\\
$^{70}$ Jeonbuk National University, Jeonju, Republic of Korea\\
$^{71}$ Johann-Wolfgang-Goethe Universit\"{a}t Frankfurt Institut f\"{u}r Informatik, Fachbereich Informatik und Mathematik, Frankfurt, Germany\\
$^{72}$ Korea Institute of Science and Technology Information, Daejeon, Republic of Korea\\
$^{73}$ KTO Karatay University, Konya, Turkey\\
$^{74}$ Laboratoire de Physique Subatomique et de Cosmologie, Universit\'{e} Grenoble-Alpes, CNRS-IN2P3, Grenoble, France\\
$^{75}$ Lawrence Berkeley National Laboratory, Berkeley, California, United States\\
$^{76}$ Lund University Department of Physics, Division of Particle Physics, Lund, Sweden\\
$^{77}$ Nagasaki Institute of Applied Science, Nagasaki, Japan\\
$^{78}$ Nara Women{'}s University (NWU), Nara, Japan\\
$^{79}$ National and Kapodistrian University of Athens, School of Science, Department of Physics , Athens, Greece\\
$^{80}$ National Centre for Nuclear Research, Warsaw, Poland\\
$^{81}$ National Institute of Science Education and Research, Homi Bhabha National Institute, Jatni, India\\
$^{82}$ National Nuclear Research Center, Baku, Azerbaijan\\
$^{83}$ National Research and Innovation Agency - BRIN, Jakarta, Indonesia\\
$^{84}$ Niels Bohr Institute, University of Copenhagen, Copenhagen, Denmark\\
$^{85}$ Nikhef, National institute for subatomic physics, Amsterdam, Netherlands\\
$^{86}$ Nuclear Physics Group, STFC Daresbury Laboratory, Daresbury, United Kingdom\\
$^{87}$ Nuclear Physics Institute of the Czech Academy of Sciences, Husinec-\v{R}e\v{z}, Czech Republic\\
$^{88}$ Oak Ridge National Laboratory, Oak Ridge, Tennessee, United States\\
$^{89}$ Ohio State University, Columbus, Ohio, United States\\
$^{90}$ Physics department, Faculty of science, University of Zagreb, Zagreb, Croatia\\
$^{91}$ Physics Department, Panjab University, Chandigarh, India\\
$^{92}$ Physics Department, University of Jammu, Jammu, India\\
$^{93}$ Physics Program and International Institute for Sustainability with Knotted Chiral Meta Matter (SKCM2), Hiroshima University, Hiroshima, Japan\\
$^{94}$ Physikalisches Institut, Eberhard-Karls-Universit\"{a}t T\"{u}bingen, T\"{u}bingen, Germany\\
$^{95}$ Physikalisches Institut, Ruprecht-Karls-Universit\"{a}t Heidelberg, Heidelberg, Germany\\
$^{96}$ Physik Department, Technische Universit\"{a}t M\"{u}nchen, Munich, Germany\\
$^{97}$ Politecnico di Bari and Sezione INFN, Bari, Italy\\
$^{98}$ Research Division and ExtreMe Matter Institute EMMI, GSI Helmholtzzentrum f\"ur Schwerionenforschung GmbH, Darmstadt, Germany\\
$^{99}$ Saga University, Saga, Japan\\
$^{100}$ Saha Institute of Nuclear Physics, Homi Bhabha National Institute, Kolkata, India\\
$^{101}$ School of Physics and Astronomy, University of Birmingham, Birmingham, United Kingdom\\
$^{102}$ Secci\'{o}n F\'{\i}sica, Departamento de Ciencias, Pontificia Universidad Cat\'{o}lica del Per\'{u}, Lima, Peru\\
$^{103}$ Stefan Meyer Institut f\"{u}r Subatomare Physik (SMI), Vienna, Austria\\
$^{104}$ SUBATECH, IMT Atlantique, Nantes Universit\'{e}, CNRS-IN2P3, Nantes, France\\
$^{105}$ Sungkyunkwan University, Suwon City, Republic of Korea\\
$^{106}$ Suranaree University of Technology, Nakhon Ratchasima, Thailand\\
$^{107}$ Technical University of Ko\v{s}ice, Ko\v{s}ice, Slovak Republic\\
$^{108}$ The Henryk Niewodniczanski Institute of Nuclear Physics, Polish Academy of Sciences, Cracow, Poland\\
$^{109}$ The University of Texas at Austin, Austin, Texas, United States\\
$^{110}$ Universidad Aut\'{o}noma de Sinaloa, Culiac\'{a}n, Mexico\\
$^{111}$ Universidade de S\~{a}o Paulo (USP), S\~{a}o Paulo, Brazil\\
$^{112}$ Universidade Estadual de Campinas (UNICAMP), Campinas, Brazil\\
$^{113}$ Universidade Federal do ABC, Santo Andre, Brazil\\
$^{114}$ Universitatea Nationala de Stiinta si Tehnologie Politehnica Bucuresti, Bucharest, Romania\\
$^{115}$ University of Cape Town, Cape Town, South Africa\\
$^{116}$ University of Derby, Derby, United Kingdom\\
$^{117}$ University of Houston, Houston, Texas, United States\\
$^{118}$ University of Jyv\"{a}skyl\"{a}, Jyv\"{a}skyl\"{a}, Finland\\
$^{119}$ University of Kansas, Lawrence, Kansas, United States\\
$^{120}$ University of Liverpool, Liverpool, United Kingdom\\
$^{121}$ University of Science and Technology of China, Hefei, China\\
$^{122}$ University of South-Eastern Norway, Kongsberg, Norway\\
$^{123}$ University of Tennessee, Knoxville, Tennessee, United States\\
$^{124}$ University of the Witwatersrand, Johannesburg, South Africa\\
$^{125}$ University of Tokyo, Tokyo, Japan\\
$^{126}$ University of Tsukuba, Tsukuba, Japan\\
$^{127}$ Universit\"{a}t M\"{u}nster, Institut f\"{u}r Kernphysik, M\"{u}nster, Germany\\
$^{128}$ Universit\'{e} Clermont Auvergne, CNRS/IN2P3, LPC, Clermont-Ferrand, France\\
$^{129}$ Universit\'{e} de Lyon, CNRS/IN2P3, Institut de Physique des 2 Infinis de Lyon, Lyon, France\\
$^{130}$ Universit\'{e} de Strasbourg, CNRS, IPHC UMR 7178, F-67000 Strasbourg, France, Strasbourg, France\\
$^{131}$ Universit\'{e} Paris-Saclay, Centre d'Etudes de Saclay (CEA), IRFU, D\'{e}partment de Physique Nucl\'{e}aire (DPhN), Saclay, France\\
$^{132}$ Universit\'{e}  Paris-Saclay, CNRS/IN2P3, IJCLab, Orsay, France\\
$^{133}$ Universit\`{a} degli Studi di Foggia, Foggia, Italy\\
$^{134}$ Universit\`{a} del Piemonte Orientale, Vercelli, Italy\\
$^{135}$ Universit\`{a} di Brescia, Brescia, Italy\\
$^{136}$ Variable Energy Cyclotron Centre, Homi Bhabha National Institute, Kolkata, India\\
$^{137}$ Warsaw University of Technology, Warsaw, Poland\\
$^{138}$ Wayne State University, Detroit, Michigan, United States\\
$^{139}$ Yale University, New Haven, Connecticut, United States\\
$^{140}$ Yonsei University, Seoul, Republic of Korea\\
$^{141}$  Zentrum  f\"{u}r Technologie und Transfer (ZTT), Worms, Germany\\
$^{142}$ Affiliated with an institute covered by a cooperation agreement with CERN\\
$^{143}$ Affiliated with an international laboratory covered by a cooperation agreement with CERN.\\

\end{flushleft} 
  
\end{document}